\pdfoutput=1

\documentclass[
    10pt,
    journal,
    letterpaper,
    twoside,
    twocolumn
]{IEEEtran}
\usepackage[T1]{fontenc}
\usepackage[utf8]{inputenc}
\usepackage[tbtags]{amsmath}
\usepackage{amssymb}
\usepackage{siunitx}
\usepackage{enumitem}
\usepackage{booktabs}
\usepackage{cite}
\usepackage{url}
\usepackage{graphicx}
\usepackage[
    subrefformat=parens,
    labelformat=parens,
    caption=false,
    font=footnotesize
]{subfig}
\usepackage[pdfborder={0 0 0}, hidelinks]{hyperref}
\usepackage[capitalize]{cleveref}
\usepackage{textcomp}
\usepackage{microtype}

\DeclareGraphicsExtensions{.pdf}
\graphicspath{{./pdf/}}

\crefformat{equation}{(#2#1#3)}
\crefrangeformat{equation}{(#3#1#4)--(#5#2#6)}
\crefname{figure}{Fig.\@}{Fig.\@}
\Crefname{figure}{Fig.\@}{Fig.\@}

\AtBeginDocument{\DeclareSIUnit{\kWh}{kWh}}
\AtBeginDocument{\DeclareSIUnit{\MWh}{MWh}}
\AtBeginDocument{\DeclareSIUnit{\kva}{kVA}}
\AtBeginDocument{\DeclareSIUnit{\mva}{MVA}}
\AtBeginDocument{\DeclareSIUnit{\pu}{pu}}
\begin{document}
\title{A Generalized PSS Architecture for Balancing\protect\\
       Transient and Small-Signal Response}
\author{%
Ryan~T.~Elliott, Payman~Arabshahi, and~Daniel~S.~Kirschen%
\thanks{R. T. Elliott, P. Arabshahi, and D. S. Kirschen are with the
Department of Electrical and Computer Engineering,
University of Washington, Seattle, WA 98195 USA
(e-mail: ryanelliott@ieee.org; paymana@uw.edu, kirschen@uw.edu).
\textit{(Corresponding author: Ryan T. Elliott.)}}%
}


\maketitle

\begin{abstract}
%
%
%
For decades, power system stabilizers paired with high initial
response automatic voltage regulators have served as an effective
means of meeting sometimes conflicting system stability requirements.
Driven primarily by increases in power electronically-coupled
generation and load, the dynamics of large-scale power systems are
rapidly changing.  Electric grids are losing inertia and traditional
sources of voltage support and oscillation damping. The system load is
becoming stiffer with respect to changes in voltage.  In parallel,
advancements in wide-area measurement technology have made it possible
to implement control strategies that act on information transmitted
over long distances in nearly real time.  In this paper, we present a
power system stabilizer architecture that can be viewed as a
generalization of the standard $\Delta\omega$-type stabilizer.  The
control strategy utilizes a real-time estimate of the
center-of-inertia speed derived from wide-area measurements.  This
approach creates a flexible set of trade-offs between transient and
small-signal response, making synchronous generators better able to
adapt to changes in system dynamics.  The phenomena of interest are
examined using a two-area test case and a reduced-order model of the
North American Western Interconnection.  To validate the key findings
under realistic conditions, we employ a state-of-the-art co-simulation
platform to combine high-fidelity power system and communication
network models.  The benefits of the proposed control strategy are
retained even under pessimistic assumptions of communication network
performance.
\end{abstract}

\begin{IEEEkeywords}
automatic voltage regulator, co-simulation, linear time-varying
systems, phasor measurement unit, power system stabilizer, real-time
control, wide-area measurement systems.
\end{IEEEkeywords}

%
\IEEEpeerreviewmaketitle

\section{Introduction}
\label{sec:intro}

\IEEEPARstart{T}{he} delicate balance between synchronizing and
damping torque components in a synchronous machine creates a
conflicting set of stability-oriented exciter performance
requirements~\cite{dandeno:68,demello:69,kundur:94,gibbard:15}.  Power
system stabilizers (PSS) have long played a critical role in
satisfying these requirements; however, changes in bulk system
dynamics pose challenges to existing control strategies.  As
inverter-coupled variable generation displaces synchronous machines,
electric grids lose inertia and traditional sources of voltage support
and oscillation damping. Correspondingly, the rapid growth of power
electronic loads may make the system load stiffer with respect to
changes in voltage~\cite{rylander:10,korunovic:18}.  In parallel with
these changes, wide-area measurement systems (WAMS) have transformed
power system monitoring.  The deployment of phasor measurement units
(PMUs) has made it is possible to implement control strategies that
act on information transmitted over long distances in nearly real
time~\cite{uhlen:12,lu:12,trudnowski:13,pierre:19}.  Despite the
proliferation of inverter-coupled resources, it is projected that
synchronous generation will account for a significant fraction of the
capacity of large-scale power systems for decades to
come~\cite{eia:19}. As the dynamics of these systems change, it may
become necessary to rethink how synchronous machines are controlled.

In this paper, we derive a new PSS architecture that can be viewed as
a generalization of the standard $\Delta\omega$-type stabilizer. This
control strategy stems from a time-varying linearization of the
equations of motion for a synchronous machine. It utilizes a real-time
estimate of the center-of-inertia speed derived from a set of
wide-area measurements. The proposed strategy improves the damping of
both local and inter-area modes of oscillation. The ability of the
stabilizer to improve damping is decoupled from its role in shaping
the system response to transient disturbances. Consequently, the
interaction between the PSS and automatic voltage regulator (AVR) can
be fine-tuned based on voltage requirements.  This approach creates a
flexible set of trade-offs between transient and small-signal
response, making synchronous generators better able to adapt to
changes in system dynamics.  Analysis and simulation show that this
strategy is tolerant of communication delay, traffic congestion, and
jitter.

\subsection{Literature Review}
\label{sec:lit_review}

The role that PSSs play in shaping the dynamic system response to
severe transient disturbances, such as generator trips, is explored
in~\cite{bollinger:79,kundur:89,grondin:93,dudgeon:07}.
In~\cite{dudgeon:07}, Dudgeon \textit{et~al.}\ show that the actions
of PSSs and AVRs are dynamically interlinked. High initial response
AVRs support transient stability but can reduce the damping of
electromechanical modes of oscillation. The primary function of PSSs
is to improve oscillation damping, but they can also degrade transient
stability by counteracting the voltage signal sent to the exciter by
the AVR.
Managing these interactions through coordinated AVR and PSS design is
studied in~\cite{law:94,quinot:99,dysko:10}.  In~\cite{grondin:93},
Grondin, Kamwa, \textit{et al.}\ present a multi-band PSS
compensator aimed at improving transient stability by adding damping
to the lowest natural resonant frequency.  The objectives of this
compensation approach are similar to those we outline in
\cref{sec:proposed_method}. We present a PSS architecture that
features a new type of multi-band compensator that leverages wide-area
measurements to achieve amplitude response attenuation.

Employing remote, or global, input signals to improve the performance
of power system damping controllers has inspired many research efforts
including~\cite{aboul:96,kamwa:01,chow:00,wu:04,zhang:13,ke:16}.
In~\cite{aboul:96}, Aboul-Ela \textit{et al.}\ propose a PSS
architecture with two inputs, a local signal used mainly for damping
the local mode and a global signal for damping inter-area modes.  For
the global signal, \cite{aboul:96} considers tie-line active power
flows and speed difference signals that provide observability of
specific inter-area modes. As stated in~\cite{grondin:93}, the ideal
stabilizing signal for a PSS ``should be in phase with the deviation
of the generator speed from the average speed of the entire system.''
To approximate this ideal signal, the rotor speed is typically passed
through a washout (highpass) filter, which may insufficiently
attenuate steady-state changes in rotor speed and/or introduce excess
phase lead into the bottom end of the control band.  In contrast, we
explore the implications of combining local measurements with a
real-time estimate of the center-of-inertia speed.

The research community is actively working to develop simulation
techniques for studying the impact of communication networks on power
systems.  \textit{Federated co-simulation environments} consist of two
or more independent simulation platforms combined so that they
exchange data and software execution commands.
In~\cite{hopkinson:06}, Hopkinson \textit{et al}.\ present EPOCHS, a
co-simulation environment that combined Network Simulator~2
(\mbox{ns-2}) with PSLF and PSCAD/EMTDC. Many subsequent research
efforts followed, including~\cite{lin:12,ciraci:14,palmintier:17}.
In this paper, the Hierarchical Engine for Large-scale Infrastructure
Co-Simulation (HELICS) is employed~\cite{palmintier:17}. We use this
state-of-the-art framework to federate a communication network model
developed in \mbox{ns-3} with a power system model developed in the
MATLAB-based Power System Toolbox (PST).

\subsection{Paper Organization}
\label{sec:paper_organization}
The remainder of this paper is organized as follows.
\Cref{sec:proposed_method} derives a generalization of the standard
$\Delta\omega$-type PSS enabled by wide-area measurements.  The impact
of this control strategy on a two-area test system is examined in
\cref{sec:two_area_analysis}.  \Cref{sec:miniwecc_analysis} evaluates
how the main results scale to large systems using a reduced-order
model of the North American Western Interconnection.  In
\cref{sec:helics_analysis}, we study the effect of nonideal
communication network performance using
co-simulation. \Cref{sec:conclusion} summarizes and concludes.

\section{Proposed Method}
\label{sec:proposed_method}
The proposed PSS architecture arises from a time-varying linearization
of the equations of motion for a synchronous machine.  Here we briefly
restate some key concepts and definitions from the theory of
continuous-time linear time-varying systems.  In the control strategy
derivation, these concepts will be applied to the nonlinear form of
the swing equation.

\subsection{Linear Time-Varying Systems}
\label{sec:ltv_systems} Let ${f:\mathbb{R}^{n} \times \mathbb{R}^{m}
\rightarrow \mathbb{R}^{n}}$ denote a nonlinear vector field
\begin{equation}
    \label{eq:nl_vector_field}
    \dot{x}(t) = f(x(t), u(t)),
\end{equation}
where ${x(t) \in \mathbb{R}^{n}}$ is the system state at time $t$ and
${u(t) \in \mathbb{R}^{m}}$ the input.  Recall that a time-varying
linearization of $f$ takes the form
\begin{equation}
    \label{eq:ltv_state_space}
    \Delta \dot{x}(t) = A(t)\Delta x(t) + B(t)\Delta u(t),
\end{equation}
where ${\Delta x(t) = x(t) - \overline{x}(t)}$ and
${\Delta u(t) = u(t) - \overline{u}(t)}$.  The time-varying trajectory
about which the system is linearized is determined by
$\overline{x}(t)$ and $\overline{u}(t)$.

The state-space matrices can be expressed compactly as
\begin{align}
    \label{eq:system_matrix}
    A(t) &= D_{x}f(\overline{x}(t),\overline{u}(t)) \\
    \label{eq:input_matrix}
    B(t) &= D_{u}f(\overline{x}(t),\overline{u}(t)),
\end{align}
where the operator $D_{x}$ returns the Jacobian matrix of partial
derivatives with respect to $x$ evaluated at time $t$, and $D_{u}$
returns the analogous matrix of partials with respect to $u$.  In
general, the state-space representation is time-varying when
$\overline{x}(t)$ and $\overline{u}(t)$ define a nonequilibrium
trajectory.

\subsection{Control Strategy Derivation}
\label{sec:strategy_derivation}

This derivation applies the concepts introduced in
\cref{sec:ltv_systems} to the equations of motion for a synchronous
machine.  Stating the nonlinear swing equation in terms of the
per-unit accelerating power, we have
\begin{equation}
    \label{eq:nl_power_swing}
    \dot{\omega}(t) = -\frac{D}{2H}\left[\omega(t) - \omega_{0}\right]
    + \frac{1}{2H\omega(t)}\left[P_m(t) - P_e(t)\right],
\end{equation}
where $\omega_{0}$ is the per-unit synchronous speed, $D$ the damping
coefficient, and $H$ the inertia constant~\cite{kundur:94,gibbard:15}.
Linearizing \cref{eq:nl_power_swing} about a nonequilibrium trajectory
yields
\begin{equation}
    \begin{split}
        \label{eq:ltv_power_swing}
        \Delta \dot{\omega}(t) ={}& {-}\left[\frac{D}{2H}
        + \frac{\overline{P}_{m}(t) - \overline{P}_{e}(t)}{2H\overline{\omega}(t)^{2}}\right]
        \Delta\omega(t) \\
        & \mspace{-\medmuskip}
        + \frac{1}{2H\overline{\omega}(t)}\left[\Delta P_{m}(t) - \Delta P_{e}(t)\right],
    \end{split}
\end{equation}
where ${\Delta \omega(t) = \omega(t) - \overline{\omega}(t)}$,
${\Delta P_{m}(t) = P_{m}(t) - \overline{P}_{m}(t)}$, and
${\Delta P_{e}(t) = P_{e}(t) - \overline{P}_{e}(t)}$.

A new damping coefficient arises from analysis of
\cref{eq:ltv_power_swing}
\begin{equation}
    \label{eq:ltv_damping}
    \mathfrak{D}(t) = D
    + \frac{\overline{P}_{m}(t) - \overline{P}_{e}(t)}{\overline{\omega}(t)^{2}}.
\end{equation}
Using this coefficient, \cref{eq:ltv_power_swing} can be restated as
\begin{equation}
    \label{eq:ltv_power_swing_ii}
    \Delta \dot{\omega}(t) = -\frac{\mathfrak{D}(t)}{2H}\Delta\omega(t)
    + \frac{1}{2H\overline{\omega}(t)}\left[\Delta P_{m}(t) - \Delta P_{e}(t)\right].
\end{equation}
Hence, as with a standard $\Delta\omega$-type stabilizer it is
possible to add damping in the LTV reference frame by creating a
component of electrical torque that is in phase with the rotor speed
deviations.  The difference is that the speed deviations are defined
such that ${\Delta \omega(t) = \omega(t) - \overline{\omega}(t)}$,
where $\overline{\omega}(t)$ is a function of time that tracks changes
in the overall system operating point. The time-varying reference
$\overline{\omega}(t)$ makes it possible to almost completely wash out
steady-state changes in rotor speed from the control error.

\subsection{Nonequilibrium Speed Trajectory}
\label{sec:speed_trajectory}

In this paper, we will examine the implications of treating
$\overline{\omega}(t)$ as a real-time estimate of the
center-of-inertia speed
\begin{equation}
    \label{eq:coi_speed}
    \overline\omega(t) \approx
    \frac{\sum_{i\in\mathcal{I}}{H_{i}\omega_{i}(t)}}{\sum_{i\in\mathcal{I}}H_{i}},
\end{equation}
where $i$ is the unit index and $\mathcal{I}$ the set of all online
conventional generators.  The right-hand side of \cref{eq:coi_speed}
corresponds to the classical center-of-inertia definition, dating back
to at least~\cite{tavora:72}. A related quantity that incorporates the
machine apparent power ratings is studied in~\cite{ulbig:14}. This
alternative approach may be more effective than \cref{eq:coi_speed} in
capturing the discrepancy in size between large and small machines
with similar inertia constants.

The question of how to compute $\overline{\omega}(t)$ for real-time
control applications is an interesting research problem in itself that
is mostly outside the scope of this paper.  A promising method is
presented in \cite{milano:18}.  At the time of this writing, rotor
speed measurements are seldom available through wide-area measurement
systems; however, a straightforward way of estimating
\cref{eq:coi_speed} would be a weighted average of frequency
measurements
\begin{equation}
    \label{eq:coi_freq}
    \overline\omega(t) = \frac{1}{f_{0}}\sum_{k\in\mathcal{K}}{\alpha_{k}f_{k}(t)},
\end{equation}
where $k$ is the sensor index, and $f_{0}$ the nominal system
frequency.  The frequency signal reported by the $k$th sensor is
denoted by $f_{k}(t)$, and the associated weight by $\alpha_{k}$.
The weights are nonnegative and sum to one, i.e., ${1^{T}\alpha = 1}$.
For simplicity, we will consider the arithmetic mean in which
${\alpha_{k} = 1/\left|\mathcal{K}\right|}$ for all $k$, where
$\left|\mathcal{K}\right|$ denotes the cardinality of $\mathcal{K}$ or
simply the number of available sensors.  The research contributions
presented in this paper do not depend strongly on this choice.  There
are numerous implementation-related issues posed by any wide-area
control scheme, such as how to handle missing or corrupted data.  For
examples of how these problems may be addressed,
see~\cite{pierre:16,pierre:19}.

\subsection{A Generalization of the $\Delta\omega$-Type PSS}
\label{sec:generalization}

\begin{figure}[!t]
    \centering
    \includegraphics[width=0.49\textwidth]{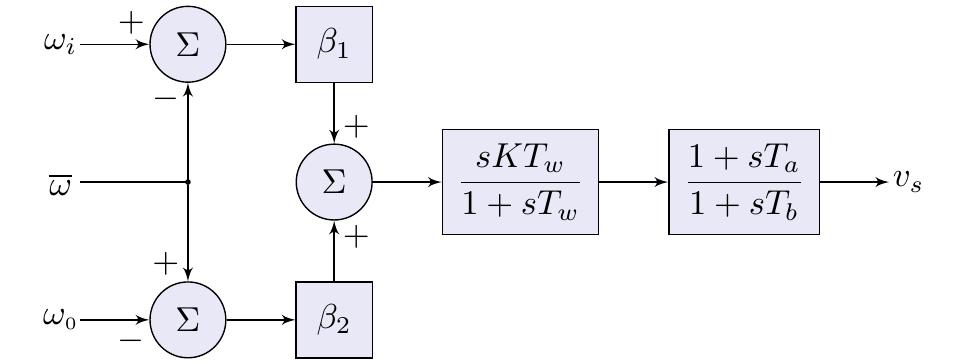}
    \caption{Generalized $\Delta\omega$-type PSS block diagram.}
    \label{fig:multi_input_pss}
\end{figure}

The control strategy implied by \cref{eq:ltv_power_swing_ii} can be
generalized to encompass the standard $\Delta\omega$-type PSS.
Splitting the linear time-invariant (LTI) control error
${\Delta\omega(t) = \omega_{i}(t)-\omega_{0}}$ into two constituent
parts and taking the linear combination yields
\begin{equation}
    \label{eq:control_error}
    \Delta\nu(t) \triangleq
    \beta_{1}\left[\omega_{i}(t) - \overline{\omega}(t)\right]
    + \beta_{2}\left[\overline{\omega}(t) - \omega_{0}\right],
\end{equation}
where $\beta_{1}$ and $\beta_{2}$ are independent tuning parameters
restricted to the unit interval.  In \cref{sec:miniwecc_analysis}, we
show how the open-loop frequency response between the input to the
exciter and the output of the PSS can be shaped by adjusting these
parameters.  The first term in~\cref{eq:control_error} follows
directly from~\cref{eq:ltv_power_swing_ii}, and the second makes it
possible to implement a standard $\Delta\omega$-type PSS using the
same framework.  As in \cref{sec:speed_trajectory},
$\overline{\omega}(t)$ is a real-time estimate of the
center-of-inertia speed.  \Cref{fig:multi_input_pss} shows the block
diagram corresponding to this control strategy where $v_{s}$ is the
output of the PSS. If necessary, more than one lead-lag compensation
stage may be employed.

The \textit{frequency regulation mode} of a power system is a very
low-frequency mode, typically below \SI{0.1}{\hertz}, in which the
rotor speeds of all synchronous generation units
participate~\cite{grondin:93,wilches:16}.  As a consequence of
synchronism, the shape of this mode is such that all conventional
generators are in phase with one another. As its name implies, the
frequency regulation mode is sensitive to load composition, turbine
governor time constants, and droop gains.  When
${\beta_{1} > \beta_{2}}$, the control tuning prioritizes the damping of
inter-area and local modes of oscillation while de-emphasizing the
frequency regulation mode. The converse is true when
${\beta_{1} < \beta_{2}}$.  In the special case that
${\beta_{1} = \beta_{2}}$, we have a conventional $\Delta\omega$-type
PSS. The resulting control error in this case is exactly the same as
in the standard formulation presented
in~\cite{kundur:94}. \Cref{tab:beta_breakdown} summarizes the effect
of $\beta_{1}$ and $\beta_{2}$ on the PSS tuning.

\begin{table}[!t]
    \renewcommand{\arraystretch}{1.15}
    \centering
    \caption{Effect of Control Parameters on PSS Tuning}
    \label{tab:beta_breakdown}
    \begin{tabular}{l l l}
        \toprule
        Parameter Values & Tuning Description \\
        \midrule
        ${\beta_{1} > \beta_{2}}$ & Targets inter-area and local modes \\
        ${\beta_{1} < \beta_{2}}$ & Targets the frequency regulation mode \\
        ${\beta_{1} = \beta_{2} \neq 0}$ & Standard $\Delta\omega$-type PSS \\
        ${\beta_{1} = \beta_{2} = 0}$ & No PSS control \\
        \bottomrule
    \end{tabular}
\end{table}

The diagram shown in \cref{fig:multi_input_pss} accurately illustrates
the control strategy; however, the structure can be clarified
further. Expanding the second term in \cref{eq:control_error} gives
\begin{equation}
    \label{eq:control_error_expanded}
    \Delta\nu(t) = \beta_{1}\left[\omega_{i}(t) - \overline{\omega}(t)\right]
    + \beta_{2}\overline{\omega}(t) - \beta_{2}\omega_{0}.
\end{equation}
Thus, we can construct the control error in~\cref{eq:control_error}
with a constant reference and a single feedback signal
\begin{align}
    \label{eq:control_error_err}
    \Delta\nu(t) &= \nu(t) - \nu_{\mathrm{ref}},\ \text{where} \\
    \label{eq:control_error_ref} \nu_{\mathrm{ref}}
    &= \beta_{2}\omega_{0},\ \text{and} \\
    \label{eq:control_error_feedback} \nu(t)
    &= \beta_{1}\left[\omega_{i}(t) - \overline{\omega}(t)\right]
    + \beta_{2}\overline{\omega}(t).
\end{align}
The results described in this paper are based on the strategy defined
by \crefrange{eq:control_error_err}{eq:control_error_feedback} and
illustrated in \cref{fig:multi_input_pss}.

\begin{figure}[!t]
    \centering
    \includegraphics[width=0.49\textwidth]{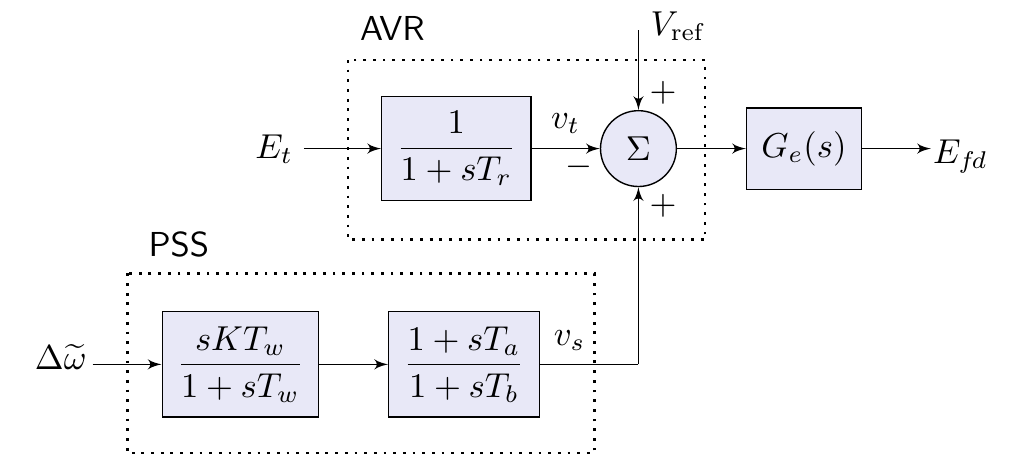}
    \caption{Excitation system with AVR and PSS, where $G_{e}(s)$
    represents the transfer function of the exciter.}
    \label{fig:avr_pss_combined}
\end{figure}

For the sake of completeness, we present a further refinement that
permits the per-unit synchronous speed $\omega_{0}$ to serve as the
reference.  The basic idea is to divide
\cref{eq:control_error_expanded} by $\beta_{2}$, taking care to
account for the case where ${\beta_{2} = 0}$.  Beginning with the
control error, we have
\begin{equation}
    \label{eq:control_error_final}
    \Delta\widetilde{\omega}(t) \triangleq
    \widetilde{\omega}(t) - \omega_{0}.
\end{equation}
The feedback signal $\widetilde{\omega}(t)$ is then given by
\begin{equation}
    \label{eq:control_error_final_feedback}
    \widetilde{\omega}(t) =
    \begin{cases}
        \left(\beta_{1}/\beta_{2}\right)
        \left[\omega_{i}(t) - \overline{\omega}(t)\right]
        + \overline{\omega}(t), &\text{for $\beta_{2} > 0$} \\[1pt]
        \beta_{1}\left[\omega_{i}(t) - \overline{\omega}(t)\right] +
        \omega_{0}, &\text{for $\beta_{2} = 0$}.
    \end{cases}
\end{equation}
This construction is similar to a conventional $\Delta\omega$-type
stabilizer where the local speed measurement $\omega_{i}(t)$ has
been replaced by $\widetilde{\omega}(t)$.  Further illustrating this
connection, when $\beta_{1}$ and $\beta_{2}$ are equal and nonzero the
feedback signal becomes ${\widetilde{\omega}(t) = \omega_{i}(t)}$.
\Cref{fig:avr_pss_combined} shows how the simplified PSS block diagram
fits in the context of an excitation system with an AVR. This form is
equivalent to the one outlined in
\crefrange{eq:control_error_err}{eq:control_error_feedback} provided
that the downstream gain $K$ is scaled appropriately.

\subsection{Comparison With Existing PSS Models}

This subsection compares the generalized $\Delta\omega$-type PSS with
two industry-standard stabilizer designs: PSS2C and PSS4C.  As
described in the IEEE recommended practice for excitation system
models~\cite{ieee:16}, PSS2C represents a flexible dual-input
stabilizer.  This model supersedes and is backward compatible with
PSS2A and PSS2B.  It may be used to represent two distinct
implementation types:
\begin{enumerate}
    \item stabilizers that utilize two inputs to estimate the
    integral of accelerating power, and
    \item stabilizers that utilize rotor speed (or frequency) feedback
    and incorporate a signal proportional to the electrical power
    as a means of compensation.
\end{enumerate}
The generalized $\Delta\omega$-type PSS presented here bears
similarities to the second of these types. The term in
\cref{eq:control_error} multiplied by $\beta_{1}$ also represents a
local rotor speed combined with an auxiliary signal.  The first key
difference is that $\overline{\omega}(t)$ in \cref{eq:control_error}
is synthesized from from wide-area, rather than strictly local,
measurements.  The second is that the generalized $\Delta\omega$ PSS
also provides the ability to independently adjust the amount of
steady-state error included in the feedback. In contrast, PSS2C
does not feature a multi-band compensation mechanism.

For multi-band compensation, we turn to PSS4C which builds upon the
structure originally proposed in~\cite{grondin:93}.  As discussed in
\cref{sec:lit_review}, the primary difference between the generalized
$\Delta\omega$ PSS and PSS4C is the way the compensation is
implemented. The PSS4C structure uses parallel lead-lag compensators
to delineate the frequency bands, whereas the generalized
$\Delta\omega$ PSS uses the linear combination of steady-state and
small-signal components in \cref{eq:control_error}.  As shown
in~\cref{sec:miniwecc_analysis}, the latter strategy provides of a
means of achieving selective attenuation with minimal impact on the
phase response.  For an in-depth comparison of PSS2B and PSS4B, the
precursors of the models discussed here, see~\cite{kamwa:05}.

\section{Two-Area System Analysis}
\label{sec:two_area_analysis}
\begin{figure}[!t]
    \centering
    \includegraphics[width=0.49\textwidth]{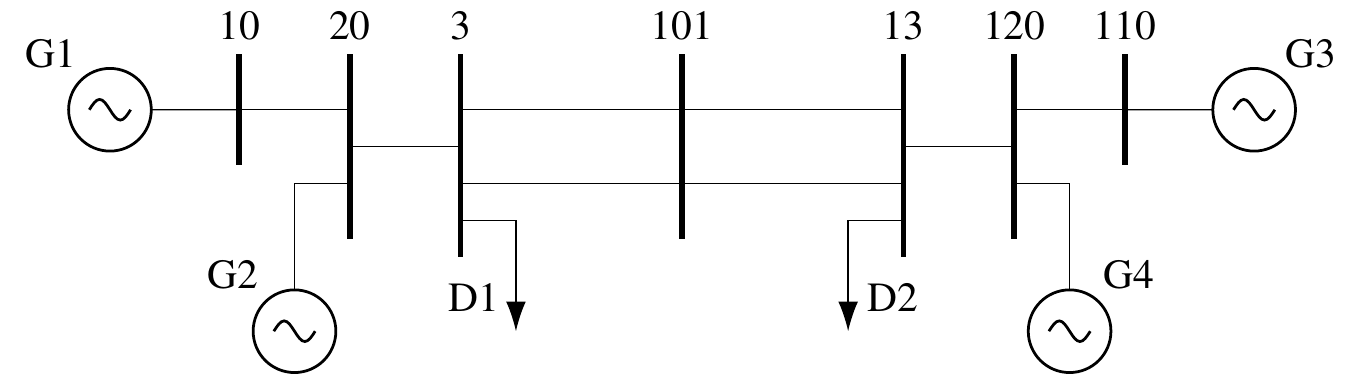}
    \caption{Oneline diagram of the two-area test system.}
    \label{fig:two_area_oneline}
\end{figure}

To study the impact of the control strategy outlined in
\cref{sec:proposed_method}, a combination of time- and
frequency-domain analysis was employed.  A custom dynamic model based
on the block diagram shown in \cref{fig:multi_input_pss} was
implemented in the MATLAB-based Power System Toolbox
(PST)~\cite{chow:92}.  This application facilitates not only
time-domain simulation of nonlinear systems but also linearization and
modal analysis.  Two test systems were studied: a small model based on
the Klein-Rogers-Kundur (KRK) two-area system~\cite{klein:91}, and a
reduced-order model of the Western Interconnection.  This section
summarizes the results of analyzing the two-area test system.  It
comprises \num{13} buses, \num{14} branches, and \num{4} synchronous
generators. A oneline diagram of the system is shown in
\cref{fig:two_area_oneline}.  In both models, the active component of
the system load is modeled as constant current and the reactive
component as constant impedance.

To permit study of transient disturbances, several modifications were
made to the original KRK system.  The synchronous machines in the
standard case are representative of aggregate groups of generators
concentrated in each area. Each unit has the same capacity and
inertia.  Hence, tripping any one generation unit offline would be
equivalent to losing \SI{25}{\percent} of the rotating inertia online
in the system.  To facilitate the study of realistically-sized
generator trips, the capacity was redistributed such that each area
possessed one machine representative of a collection of generators and
the other a large individual plant.  Generators G\num{1} and G\num{3}
were scaled such that they each represented \SI{5}{\percent} of the
overall system capacity. The remainder was equally split between
G\num{2} and G\num{4}.  Every unit in the system was then outfitted
with the generalized $\Delta\omega$ PSS described in
\cref{sec:proposed_method}.

\subsection{Sensitivity of System Poles to the PSS Tuning Parameters}
\label{sec:two_area_poles}

\begin{figure}[!t]
    \captionsetup[subfloat]{farskip=0pt}
    \centering
    \subfloat[$\beta_{1}$ parameter sweep.]{
        \centering
        \includegraphics[width=0.232\textwidth]{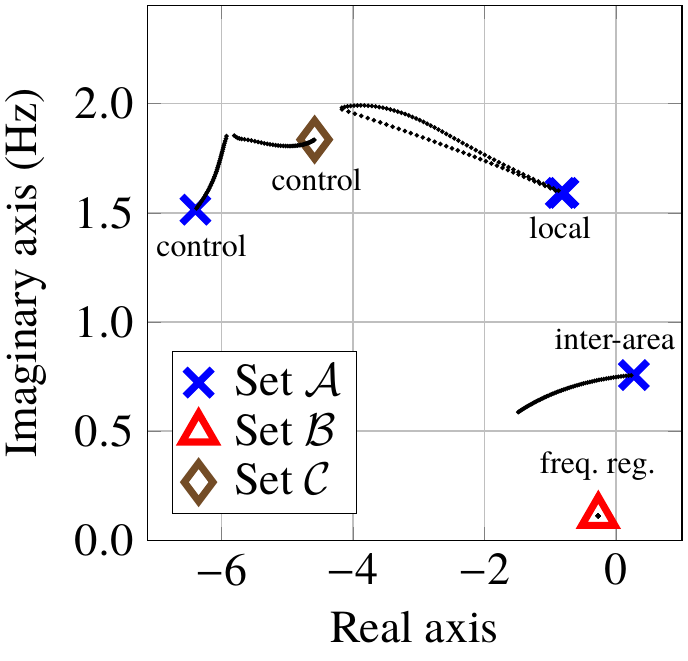}
        \label{fig:krk_beta1_sensitivity}
    }\thinspace
    \subfloat[$\beta_{2}$ parameter sweep.]{
        \centering
        \includegraphics[width=0.232\textwidth]{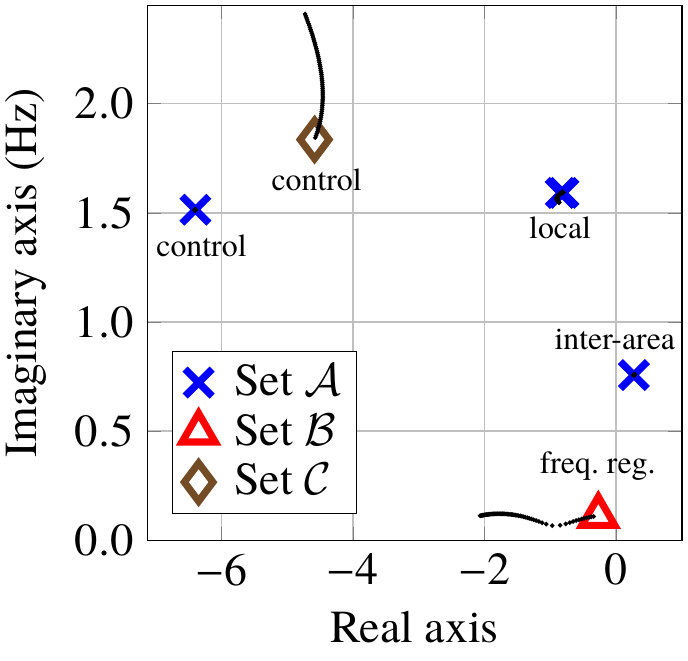}
        \label{fig:krk_beta2_sensitivity}
    }
    \caption{Sensitivity of the system oscillatory modes to the PSS
    tuning parameters.  The modes in $\mathcal{A}$ are sensitive to
    changes in $\beta_{1}$, those in $\mathcal{B}$ to $\beta_{2}$, and
    those in $\mathcal{C}$ to both.}
    \label{fig:krk_pole_sensitivity}
\end{figure}

\begin{figure}[!t]
    \captionsetup[subfloat]{farskip=0pt}
    \centering
    \subfloat[\SI{0.76}{\hertz} inter-area mode.]{
        \centering
        \includegraphics[width=0.232\textwidth]{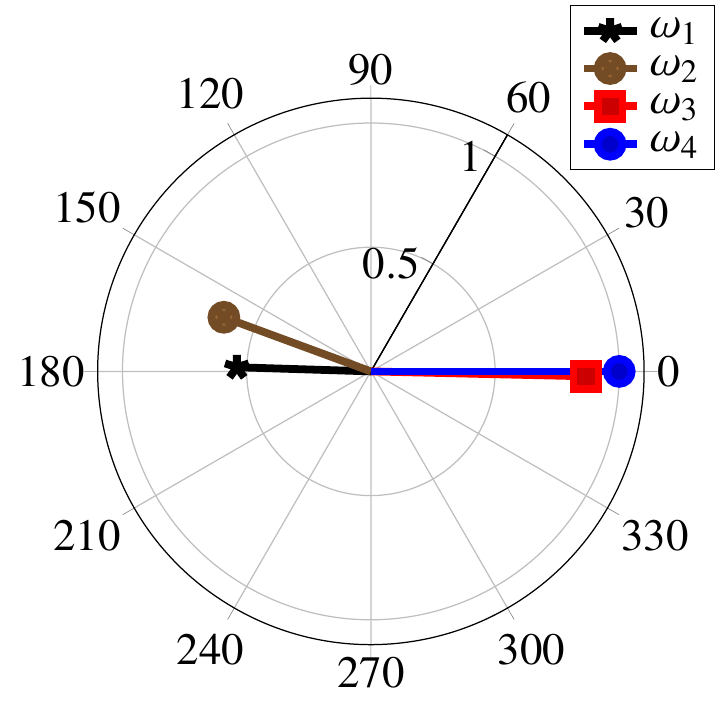}
        \label{fig:mode_shape_inter}
    }\thinspace
    \subfloat[\SI{0.09}{\hertz} freq.\ regulation mode.]{
        \centering
        \includegraphics[width=0.232\textwidth]{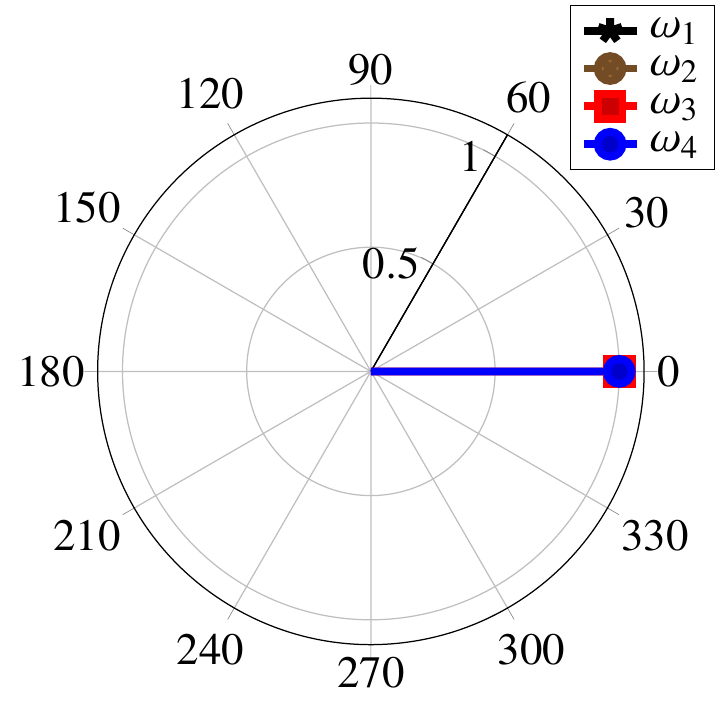}
        \label{fig:mode_shape_freq}
    }
    \caption{Normalized mode shape plots for the two-area system.}
    \label{fig:mode_shape}
\end{figure}

Here we examine the effects of sweeping the PSS tuning parameters
$\beta_{1}$ and $\beta_{2}$ on the poles of the system. The modal
analysis was performed by linearizing the system dynamics and then
solving for the eigenvalues and eigenvectors of the system matrix.
The main result is that the oscillatory modes effectively split into
two groups, one that is sensitive to changes in $\beta_{1}$ and the
other $\beta_{2}$.  Let us begin by examining the effect of the tuning
parameters on the inter-area and local modes.  Consider the inter-area
mode indicated by the blue x located at \SI{0.76}{\hertz} in
\cref{fig:krk_pole_sensitivity}.  The shape of this mode observed
through the machine speeds is shown in
Fig.~\subref*{fig:mode_shape_inter}.  Recall that mode shape is
defined by the elements of the right eigenvector corresponding to the
states of interest~\cite{rogers:99}. As demonstrated by
Fig.~\subref*{fig:mode_shape_inter}, this mode is characterized by
generators G1 and G2 oscillating against G3 and G4. The two-area
system is tuned such that this inter-area mode is unstable without
supplemental damping control.

The plots in \cref{fig:krk_pole_sensitivity} show the sensitivity of
the system poles to the PSS tuning parameters.  To generate these
plots, either $\beta_{1}$ or $\beta_{2}$ was swept over an interval
while the other was held at zero.  The tuning parameters for all of
the PSS units were swept in unison, and the gain was uniformly held
fixed at ${K = 25}$.  Sweeping the tuning parameters for all units
together facilitates study of the effect of PSSs on the frequency
regulation mode. In Fig.~\subref*{fig:krk_beta1_sensitivity},
$\beta_{1}$ was swept over the interval $[0,1]$ while $\beta_{2}$ was
held at zero.  As $\beta_{1}$ increases, the inter-area mode moves to
the left and decreases slightly in frequency. The local modes,
indicated by the blue x's in the upper right quadrant of
Fig.~\subref*{fig:krk_beta1_sensitivity}, move to the left and
increase slightly in frequency. A well-controlled exciter mode marked
by the blue x in the upper left quadrant moves up and to the right but
remains comfortably in the left half of the complex plane.  For all
intents and purposes, the frequency regulation mode is unaffected by
changes in $\beta_{1}$. Hence, $\beta_{1}$ dictates the extent to
which the PSS damps inter-area and local modes of oscillation.

The parameter $\beta_{2}$ primarily influences the frequency
regulation mode.  This mode is indicated by the red triangle located
at \SI{0.09}{\hertz} in \cref{fig:krk_pole_sensitivity}.  The shape of
the frequency regulation mode observed through the machine speeds is
shown in Fig.~\subref*{fig:mode_shape_freq}.  All of the machine
speeds are in phase and have nearly identical magnitudes.  In
Fig.~\subref*{fig:krk_beta2_sensitivity}, the parameter $\beta_{2}$
was swept over the interval $[0,1]$ while $\beta_{1}$ was held at
zero.  As $\beta_{2}$ increases, the frequency regulation mode moves
to the left.  The higher-frequency exciter mode marked with a diamond
moves upward.  This control mode exhibits some sensitivity to both
$\beta_{1}$ and $\beta_{2}$.  In contrast, the inter-area and local
modes are relatively unaffected by changes in $\beta_{2}$.  The
dependence of the frequency regulation mode on $\beta_{2}$ indicates
that PSSs, in aggregate, play an important role in shaping the system
response to transient disturbances. To demonstrate this phenomenon,
and the effects of the PSS tuning parameters more broadly, we present
a collection of time-domain simulations.

\subsection{Time-Domain Simulations}
\label{sec:two_area_time}
The two-area system was simulated in PST for a variety of PSS
tunings. As in the frequency-domain analysis, all PSS units were tuned
alike and used the same gain.  The contingency of interest in this set
of simulations is a trip of generator G3. This event was selected
because it initiates a transient disturbance that excites not only the
inter-area and local modes but also the frequency regulation mode.  In
the first set of simulations, $\beta_{1}$ was varied over the set
$\{0.33,0.67,1\}$ while $\beta_{2}$ was held fixed at $0.33$. In the
second set of simulations, $\beta_{1}$ was held fixed at $0.33$ while
$\beta_{2}$ was varied over the set $\{0,0.33,0.67\}$. For all
simulations, the overall PSS gain was set to ${K = 18}$. The case
where ${\beta_{1} = \beta_{2} = 0.33}$ corresponds to a standard
$\Delta\omega$ stabilizer with a gain of ${K = 6}$.  This set of
simulations assumes ideal communication in the construction of the
time-varying reference $\overline{\omega}(t)$.
\Cref{sec:helics_analysis} addresses the effect of nonideal
communication network performance.
%

\begin{figure}[!t]
    \centering
    \includegraphics[width=0.395\textwidth]{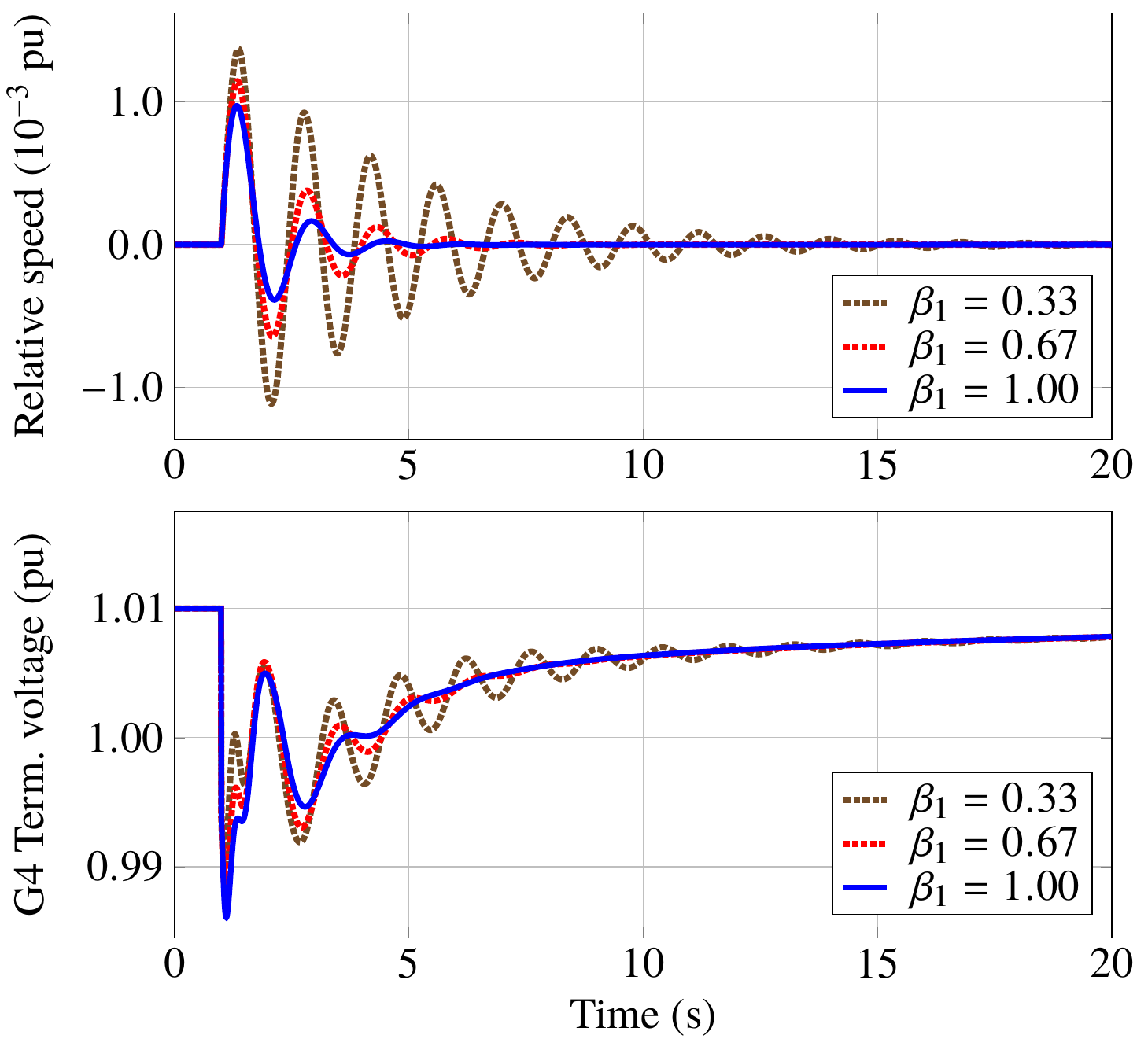}
    \caption{Time-domain simulations of generator G3 being tripped
    offline for various values of $\beta_{1}$. The top subplot shows the
    relative speed between G2 and G4.}
    \label{fig:beta1_time_plots}
\end{figure}

\Cref{fig:beta1_time_plots} shows the key results for the case where
$\beta_{1}$ is varied. The upper subplot shows the difference in speed
between generators G2 and G4, i.e., ${\omega_{2}(t)-\omega_{4}(t)}$.
The oscillatory content in this signal is dominated by the
\SI{0.76}{\hertz} inter-area mode.  As $\beta_{1}$ increases, the
damping of this mode also increases.  The lower subplot shows the
terminal voltage of generator G4. As $\beta_{1}$ is varied, the
large-signal trajectory of the terminal voltage and its
post-disturbance value are unchanged.  This reflects the fact that
varying $\beta_{1}$ only alters the small-signal characteristics of
the field current.

\begin{figure}[!t]
    \centering
    \includegraphics[width=0.395\textwidth]{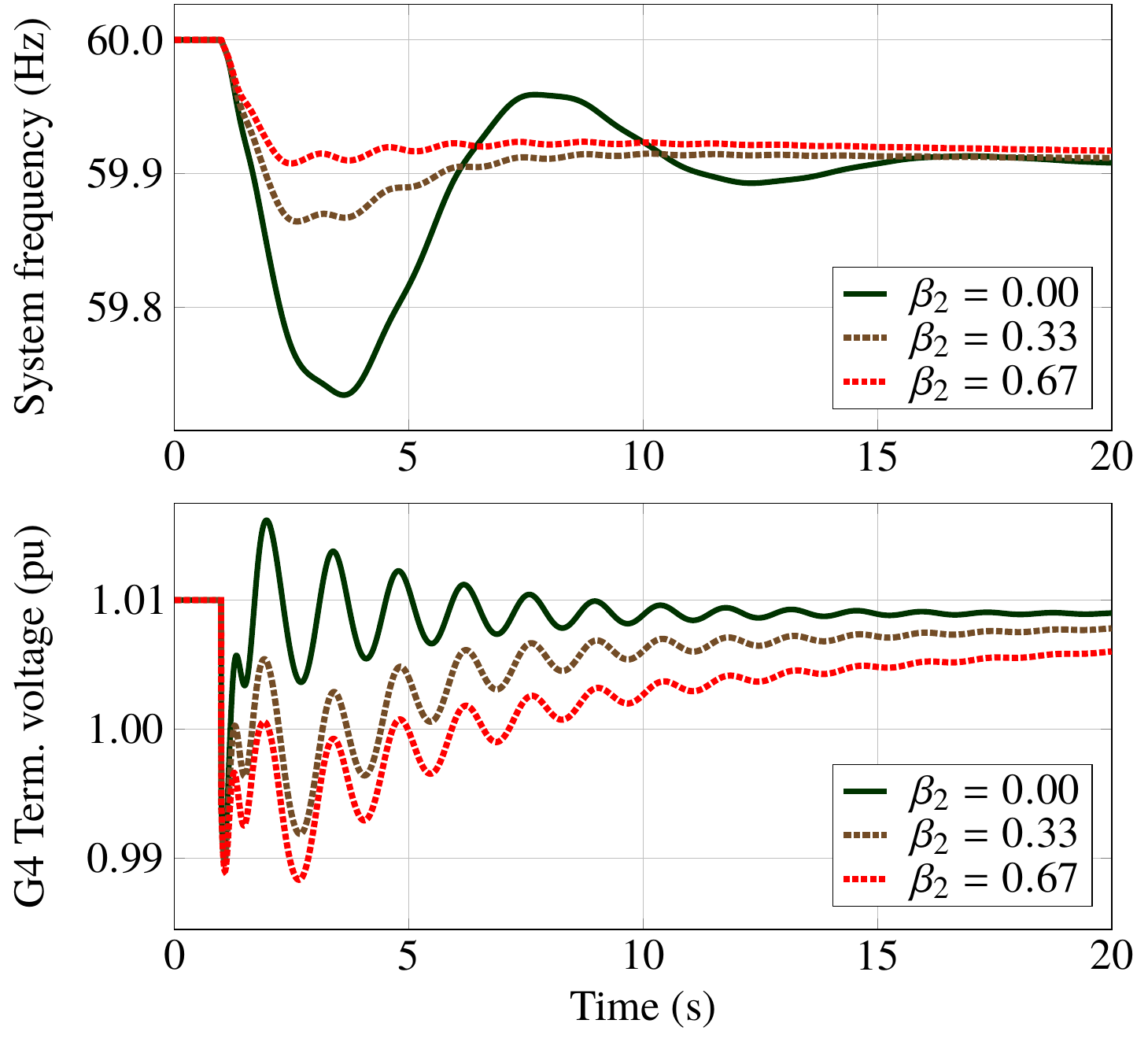}
    \caption{Time-domain simulations of generator G3 being tripped
    offline for various values of $\beta_{2}$.}
    \label{fig:beta2_time_plots}
\end{figure}

\Cref{fig:beta2_time_plots} shows the key results for the case where
$\beta_{2}$ is varied. The upper subplot shows the system frequency
response, which readily shows the behavior of the frequency regulation
mode. The results show that $\beta_{2}$ plays a key role in
determining the depth of the frequency nadir.  The frequency nadir
improves significantly as $\beta_{2}$ increases from \num{0} to
\num{0.33}, and modestly as it goes from \num{0.33} to \num{0.67}.
Effectively, $\beta_{2}$ determines the level of overshoot in the
system step response. The lower subplot shows the terminal voltage of
generator G4. As $\beta_{2}$ is increased, the terminal voltage
following the generator trip becomes incrementally more depressed.
This can be attributed to the fact that $\beta_{2}$ controls the
extent to which steady-state changes in rotor speed are included in
the PSS control error.

The process that causes $\beta_{2}$ to affect the frequency nadir is
indirect. Increasing $\beta_{2}$ amplifies the steady-state component
of the control error in \cref{eq:control_error}. This depresses the
field current supplied by the exciter and causes the voltage induced
in the stator to dip. The electrical load decreases in response to
this voltage dip with the amount of relief depending on the
sensitivity of the load with respect to voltage.  This tends to reduce
the time-varying mismatch in mechanical and electrical torque, which
improves the frequency nadir.  This effect depends on the load
composition, and the amount of improvement in the nadir decreases as
the fraction of constant power load increases.  Thus, there is a
trade-off between improving the frequency nadir and degrading the
voltage response.  As explained in~\cite{dudgeon:07}, the tendency of
the PSS to counteract the voltage signal sent to the exciter by the
AVR can reduce synchronizing torque and degrade transient stability.
The control strategy presented in this paper makes it possible to
fine-tune the interaction between the PSS and AVR without affecting
the damping of inter-area and local modes, and vice versa.
%
\section{Large-Scale Test System Analysis}
\label{sec:miniwecc_analysis}
For the two-area system discussed in \cref{sec:two_area_analysis}, the
inter-area and local modes were influenced by $\beta_{1}$, and the
frequency regulation mode by $\beta_{2}$.  This section addresses
whether this property is preserved for large-scale systems.  We
consider a reduced-order model of the Western Interconnection named
the \textit{miniWECC}, in reference to the Western Electric
Coordinating Council (WECC). It comprises \num{122} buses, \num{171}
ac branches, \num{2} HVDC lines, and \num{34} synchronous generators.
This system spans the entirety of the interconnection including
British Columbia and Alberta.  Its modal properties have been
extensively validated against real system
data~\cite{trudnowski:13,trudnowski:14}.  The aim of this analysis is
to illustrate the fundamental behavior of various aspects of the
proposed architecture in a controlled setting. Prior to
implementation, high-fidelity simulation studies that account for
variation in PSS structure and the dynamics of inverter-coupled
generation would be required.

\subsection{Sensitivity of System Poles to the PSS Tuning Parameters}
\label{sec:mini_sensitivity}
\begin{figure}[!t]
    \captionsetup[subfloat]{farskip=0pt}
    \centering
    \subfloat[$\beta_{1}$ parameter sweep.]{
        \centering
        \includegraphics[width=0.232\textwidth]{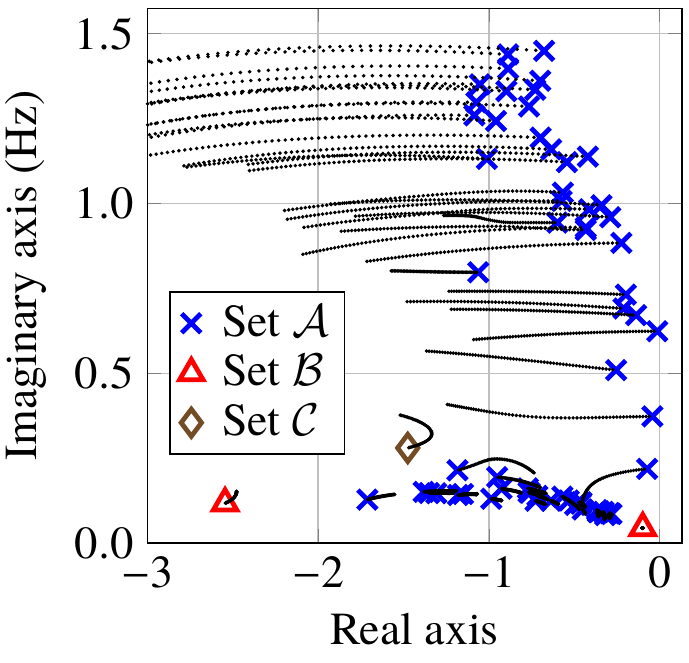}
        \label{fig:mini_beta1_sensitivity}
    }\thinspace
    \subfloat[$\beta_{2}$ parameter sweep.]{
        \centering
        \includegraphics[width=0.232\textwidth]{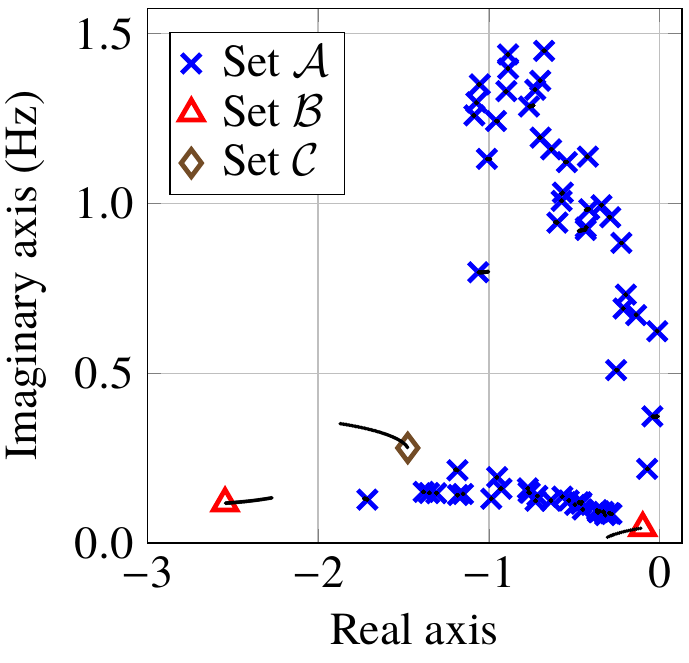}
        \label{fig:mini_beta2_sensitivity}
    }
    \caption{Sensitivity of the miniWECC oscillatory modes to the PSS
    tuning parameters.  The modes in $\mathcal{A}$ are sensitive to
    changes in $\beta_{1}$, those in $\mathcal{B}$ to $\beta_{2}$, and
    those in $\mathcal{C}$ to both.}
    \label{fig:mini_pole_sensitivity}
\end{figure}

To examine the sensitivity of the oscillatory modes to the PSS tuning
parameters, the method described in~\cref{sec:two_area_poles} was
applied to the miniWECC. Every generation unit in the system was
outfitted with a generalized $\Delta\omega$ PSS with the gain set to
${K = 25}$.  In practice, WECC policy dictates that ``a PSS shall be
installed on every synchronous generator that is larger than
\SI{30}{\mva}, or is part of a complex that has an aggregate capacity
larger than \SI{75}{\mva}, and is equipped with a suitable excitation
system''~\cite{wecc:02}.  \Cref{fig:mini_pole_sensitivity} shows the
movement of the system poles in response to changes in the tuning
parameters.  In each subplot, either $\beta_{1}$ or $\beta_{2}$ was
swept over an interval while the other was held at zero. The main
result matches the one observed for the two-area system.  The
inter-area and local modes are influenced by $\beta_{1}$, and the
frequency regulation mode by $\beta_{2}$.  For the miniWECC, there is
one well-controlled exciter mode near \SI{0.28}{\hertz} that exhibits
sensitivity to both parameters. This mode is marked in
\cref{fig:mini_pole_sensitivity} with a diamond.

\subsection{Open-Loop Frequency Response Analysis}
\label{sec:loop_response}

The frequency-domain analysis presented in
Sections~\ref{sec:two_area_analysis} and~\ref{sec:mini_sensitivity}
focused on a system-wide perspective.  Here we provide a unit-specific
analysis of the open-loop frequency response for a single
generator. Outfitting a single unit with a PSS yields the state-space
representation
\begin{align}
    \label{eq:lti_state_space}
    \dot{x}(t) &= Ax(t) + B_{p}u(t) \\
    \label{eq:lti_output}
    y_{\nu}(t) &= C_{\nu}x(t),
\end{align}
where $B_{p}$ describes how the system states are affected by changes
in the PSS control input.  The closed-loop control action determined
by~\cref{eq:control_error_err} can be implemented with the input
\begin{align}
    \label{eq:lti_input_short}
    u(t) &= -Ky_{\nu}(t) = -KC_{\nu}x(t) \\
    \label{eq:lti_input_long}
    u(t) &= -K
        \begin{bmatrix}
            0 & \gamma_{1} & \gamma_{2} & \dots & -\beta_{1}
        \end{bmatrix}
        \begin{bmatrix}
            \widehat{x}(t) \\ f_{1}(t) \\ f_{2}(t) \\ \vdots
            \\ \omega_{i}(t)
        \end{bmatrix},
\end{align}
where $K$ is a scalar gain. The output matrix $C_{\nu}$ combines the
states to form the PSS feedback signal $\nu(t)$. Note the presence of
the extra negative sign to conform to the negative feedback
convention. The state vector $x$ in \cref{eq:lti_input_long} is
organized with the unused states $\widehat{x}$ on top, followed by the
frequency measurements and the local rotor speed. For the $k$th sensor
${\gamma_{k} = \alpha_{k}\left(\beta_{1} - \beta_{2}\right)/f_{0}}$,
where $\alpha_{k}$ stems from the linear combination
in~\cref{eq:coi_freq}, and $f_{0}$ is the nominal system frequency.
In this analysis, the frequencies were computed by applying a
derivative-filter cascade to the bus voltage angles as described
in~\cite{pierre:19}.  Hence, the unity-gain open-loop transfer
function between a change in the PSS reference $\nu_{\mathrm{ref}}$
and the feedback signal $\nu$ is
\begin{equation}
    \label{eq:loop_xfer_fun}
    H(s) = C_{\nu}(sI - A)^{-1}B_{p}.
\end{equation}

\begin{figure}[!t]
    \centering
    \includegraphics[width=0.44\textwidth]{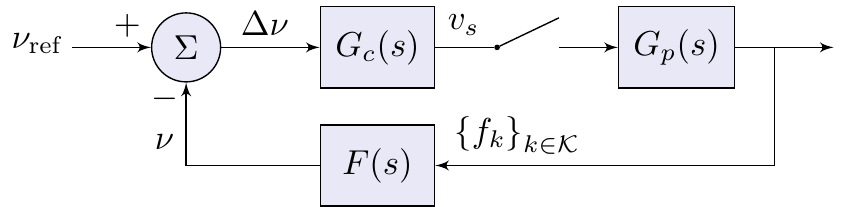}
    \caption{Feedback loop for a single generation unit outfitted with
    a generalized $\Delta\omega$ PSS, where $G_{c}(s)$ represents the
    controller, $G_{p}(s)$ the plant, and $F(s)$ the feedback process.}
    \label{fig:feedback_loop}
\end{figure}

\Cref{fig:feedback_loop} shows a high-level block diagram of the
feedback loop for a single generation unit outfitted with a
generalized $\Delta\omega$ PSS.  Here $G_{c}(s)$ represents the PSS,
$G_{p}(s)$ the plant, and $F(s)$ the feedback process. The exciter
dynamics are included in $G_{p}(s)$, and the input to the plant
represents a change in the exciter voltage reference
$V_{\mathrm{ref}}$.  By commutativity, it holds that
\begin{equation}
    H(s) = G_{c}(s)G_{p}(s)F(s) = G_{p}(s)F(s)G_{c}(s).
\end{equation}
Hence, the loop transfer function between $\Delta\nu$ and $\nu$ is the
same as the transfer function between a change in the exciter voltage
reference $V_{\mathrm{ref}}$ and the output of the PSS $v_{s}$.  Using
this function, we can evaluate the effect of the PSS tuning parameters
on the open-loop frequency response.  For this analysis, only the unit
being studied was outfitted with a PSS.

\begin{figure}[!t]
    \centering
    \includegraphics[width=0.395\textwidth]{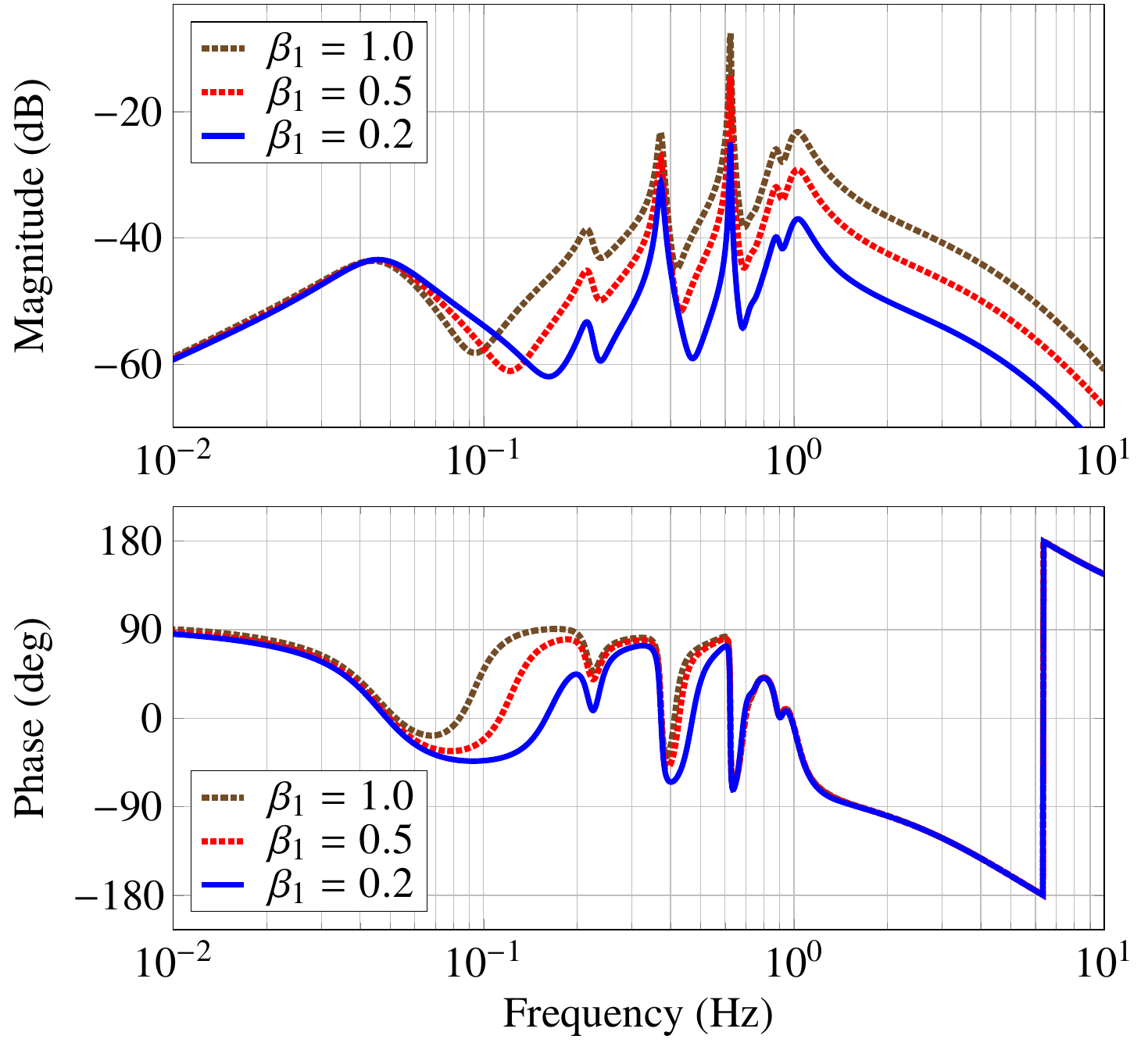}
    \caption{The effect of $\beta_{1}$ on the open-loop frequency response
    between the input to the exciter and the output of the generalized
    PSS for generator G2.}
    \label{fig:beta1_xfer_fun}
\end{figure}

\Cref{fig:beta1_xfer_fun} shows the effect of $\beta_{1}$ on the
open-loop frequency response for generator G2, a hydroelectric unit in
eastern British Columbia, where ${\beta_{2} = 1}$ for all traces.  The
peak in the amplitude response near \SI{0.04}{\hertz} corresponds to
the frequency regulation mode.  As the plot shows, $\beta_{1}$ has no
effect on the gain of the system at this frequency.  This corroborates
the system-wide modal analysis done in
Sections~\ref{sec:two_area_analysis} and~\ref{sec:mini_sensitivity} at
the unit level. As expected, $\beta_{1}$ does change the amplitude
response for the inter-area and local modes of oscillation.  Unlike
traditional compensation methods, this approach does not degrade the
phase response in the attenuation region.  As $\beta_{1}$ is varied,
the phase response at the frequencies of the dominant amplitude peaks
(\SI{0.37}{\hertz}, \SI{0.62}{\hertz}, and \SI{1.0}{\hertz}) is
essentially unchanged. The observed transition in phase through
\SI{0}{\degree} at the resonant frequencies is ideal for damping
control.

\begin{figure}[!t]
    \centering
    \includegraphics[width=0.395\textwidth]{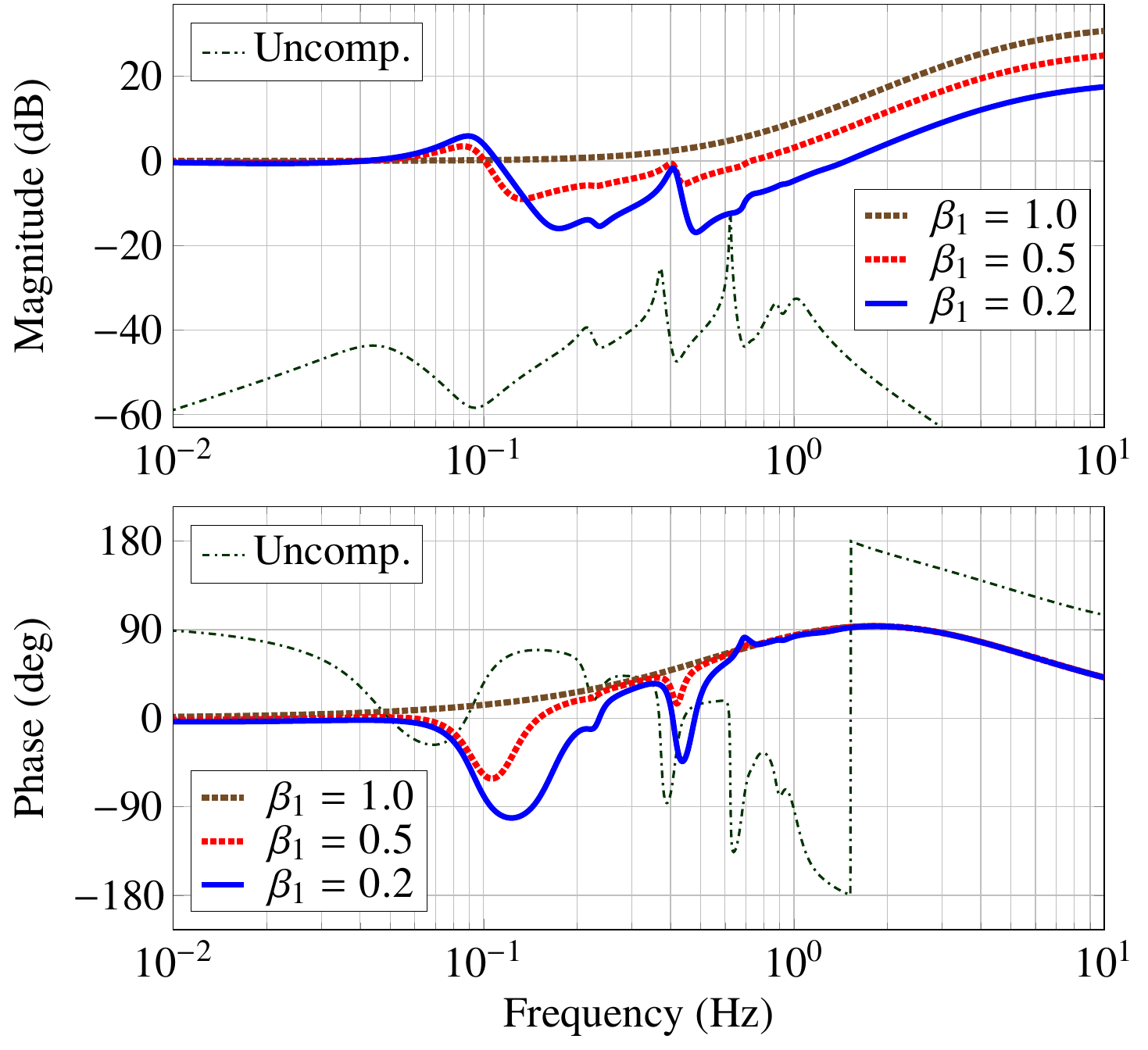}
    \caption{The effect of $\beta_{1}$ on the overall PSS compensation for
    generator G2 with the washout filter included in the uncompensated
    frequency response.}
    \label{fig:beta1_xfer_fun_uncomp}
\end{figure}

\Cref{fig:beta1_xfer_fun_uncomp} shows the effect of $\beta_{1}$ on
the overall PSS compensation.  As in~\cite{tutor:09}, the
uncompensated open-loop frequency response, including the washout
filter dynamics, is provided for comparison.  The overall compensation
comprises both the lead-lag compensator and the tuning determined by
${\beta_{1},\,\beta_{2}}$.  When ${\beta_{1} = \beta_{2} = 1}$, the tuning
stage has a gain of unity and imparts no phase shift. Hence, all of
the compensation stems from the lead-lag compensator. This is expected
because the case where ${\beta_{1} = \beta_{2} = 1}$ yields a standard
$\Delta\omega$ stabilizer as shown in \cref{tab:beta_breakdown}.

\begin{figure}[!t]
    \centering
    \includegraphics[width=0.395\textwidth]{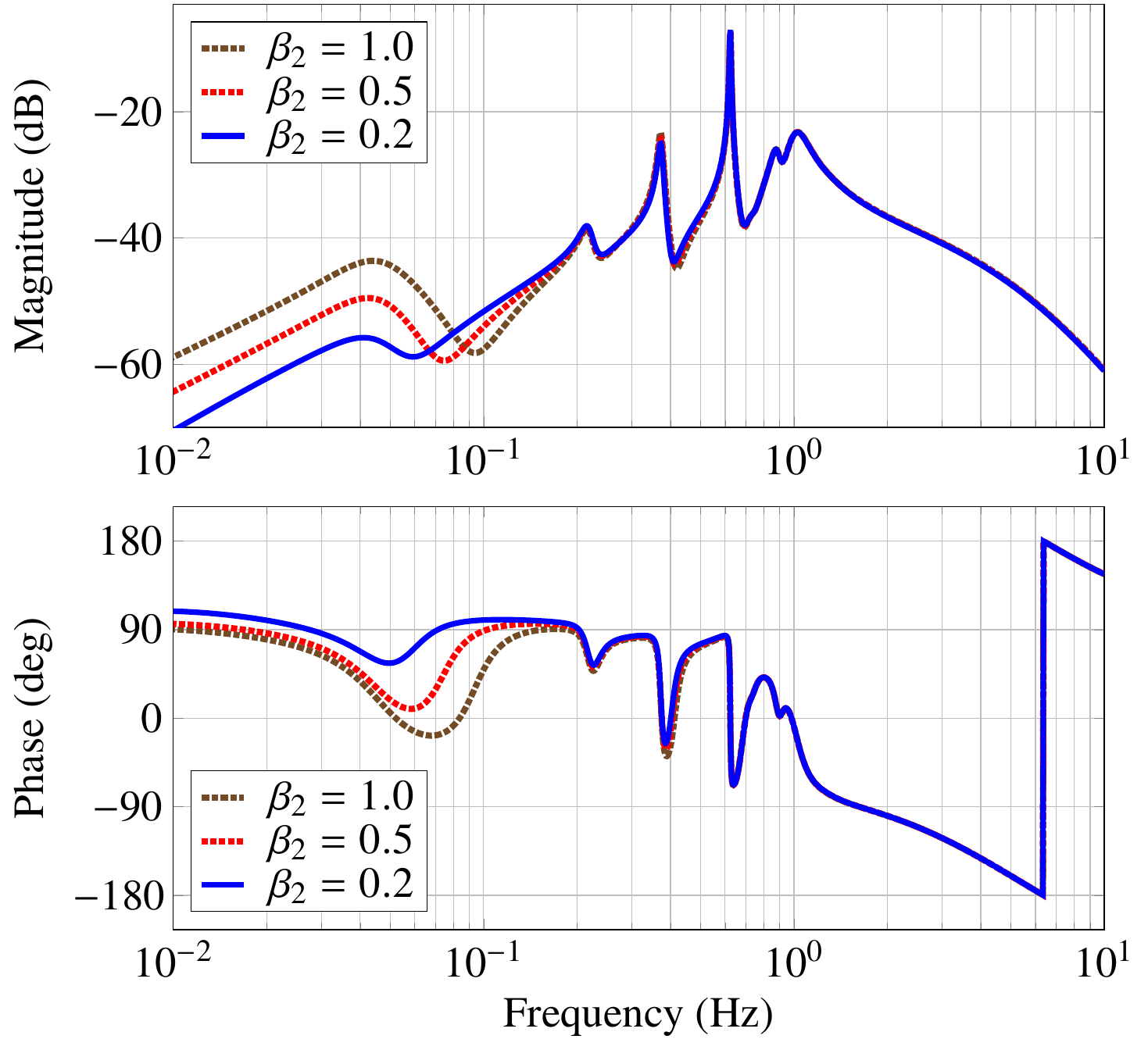}
    \caption{The effect of $\beta_{2}$ on the open-loop frequency response
    between the input to the exciter and the output of the generalized
    PSS for generator G2.}
    \label{fig:beta2_xfer_fun}
\end{figure}

\begin{figure}[!t]
    \centering
    \includegraphics[width=0.395\textwidth]{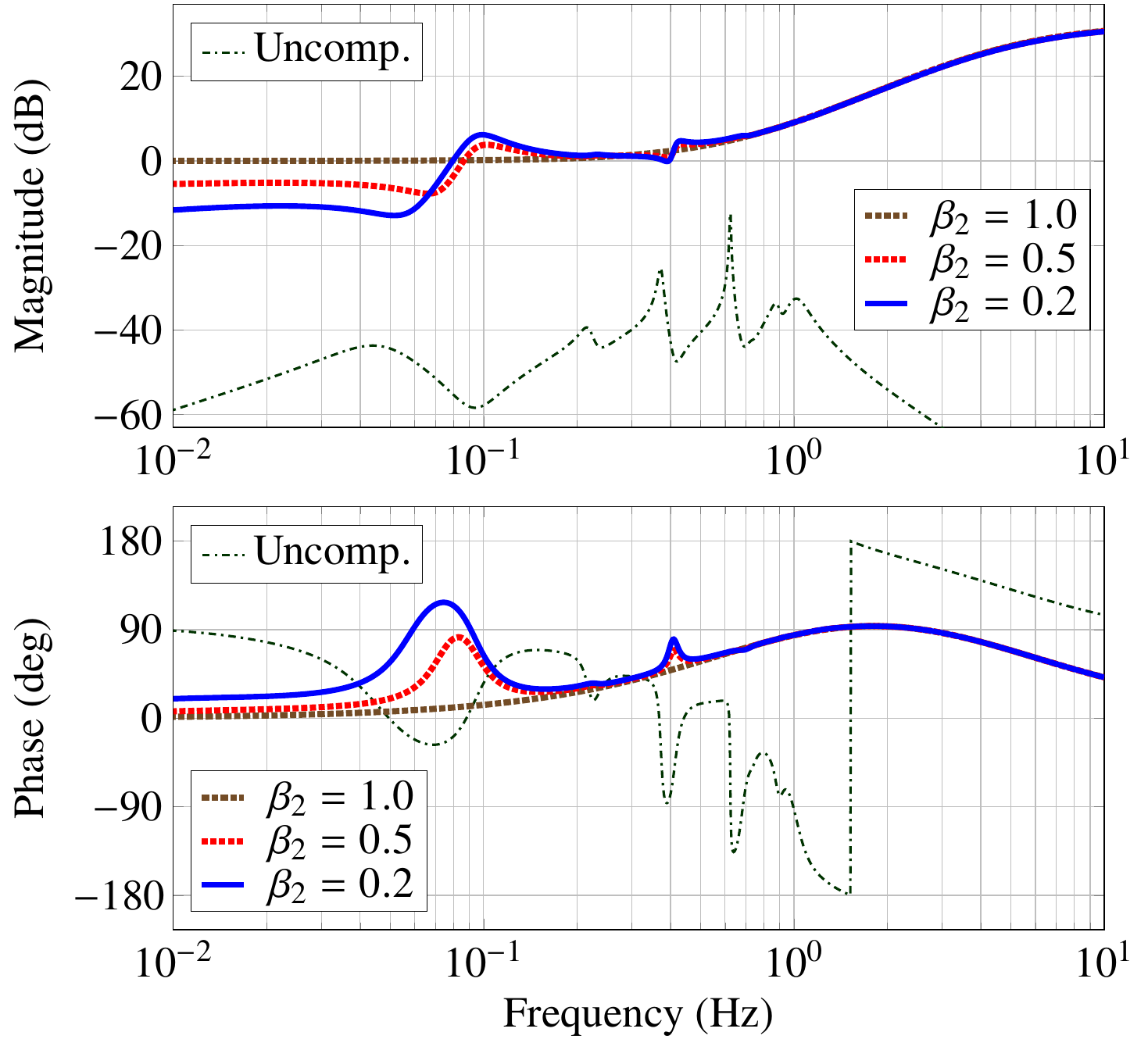}
    \caption{The effect of $\beta_{2}$ on the overall PSS compensation for
    generator G2 with the washout filter included in the uncompensated
    frequency response.}
    \label{fig:beta2_xfer_fun_uncomp}
\end{figure}

\Cref{fig:beta2_xfer_fun} shows the effect of $\beta_{2}$ on the
open-loop frequency response where ${\beta_{1} = 1}$ for all traces.
As $\beta_{2}$ is varied, the amplitude response at the frequencies
corresponding to the local and inter-area modes is effectively
unchanged. In contrast, the gain at the frequency regulation mode is
reduced by roughly \SI{14}{\decibel} as $\beta_{2}$ goes from \num{1}
to \num{0.2}.  For ${\beta_{2} = 0.2}$, the phase response at the
frequency regulation mode leads the case where ${\beta_{2} = 1}$ by
roughly \SI{35}{\degree}.  This suggests that if a $\beta_{2}$ value
below some nominal threshold is required for a particular application,
it may be necessary to retune the lead-lag compensator and/or washout
filter to ensure satisfactory low-frequency performance.  The effect
of $\beta_{2}$ on the overall PSS compensation is shown in
\cref{fig:beta2_xfer_fun_uncomp}.

\subsection{Co-Simulation of Power and Communication Systems}
\label{sec:helics_analysis}

All of the analysis presented in Sections~\ref{sec:two_area_analysis}
through \ref{sec:loop_response} was performed under the assumption of
ideal communication. In this section, we analyze the effect of
communication delay in the frequency domain and verify the findings in
the time domain, as in~\cite{wilches:17}.  The mathematical modeling
developed here represents the real-time exchange of synchronized
phasor measurement data over a network.  As described in the IEEE
standard governing data transfer in PMU networks~\cite{ieee:11},
communication delays in WAMS are typically in the range of
\num{20}--\SI{50}{\milli\second}; however, the combined delay must
also account for the effect of transducers, processing, concentrators,
and multiplexing~\cite{nad:02,zhang:14,wilches:def}.
In~\cite{nad:02}, the delay attributed to these factors is estimated
at~\SI{75}{\milli\second}, which yields an approximate range of
\num{95}--\SI{125}{\milli\second} for the combined delay. This range
is reflective of systems that utilize fiber-optic communication. It
aligns closely with the experimental results reported
in~\cite{pierre:19}, \num{69}--\SI{113}{\milli\second}, but may vary
depending on the communication method employed, e.g., wired vs.\
wireless.  Here we evaluate scenarios with delays that are \num{5} to
\num{10} times greater than the high end of this range.

\begin{figure}[!t]
    \centering
    \includegraphics[width=0.395\textwidth]{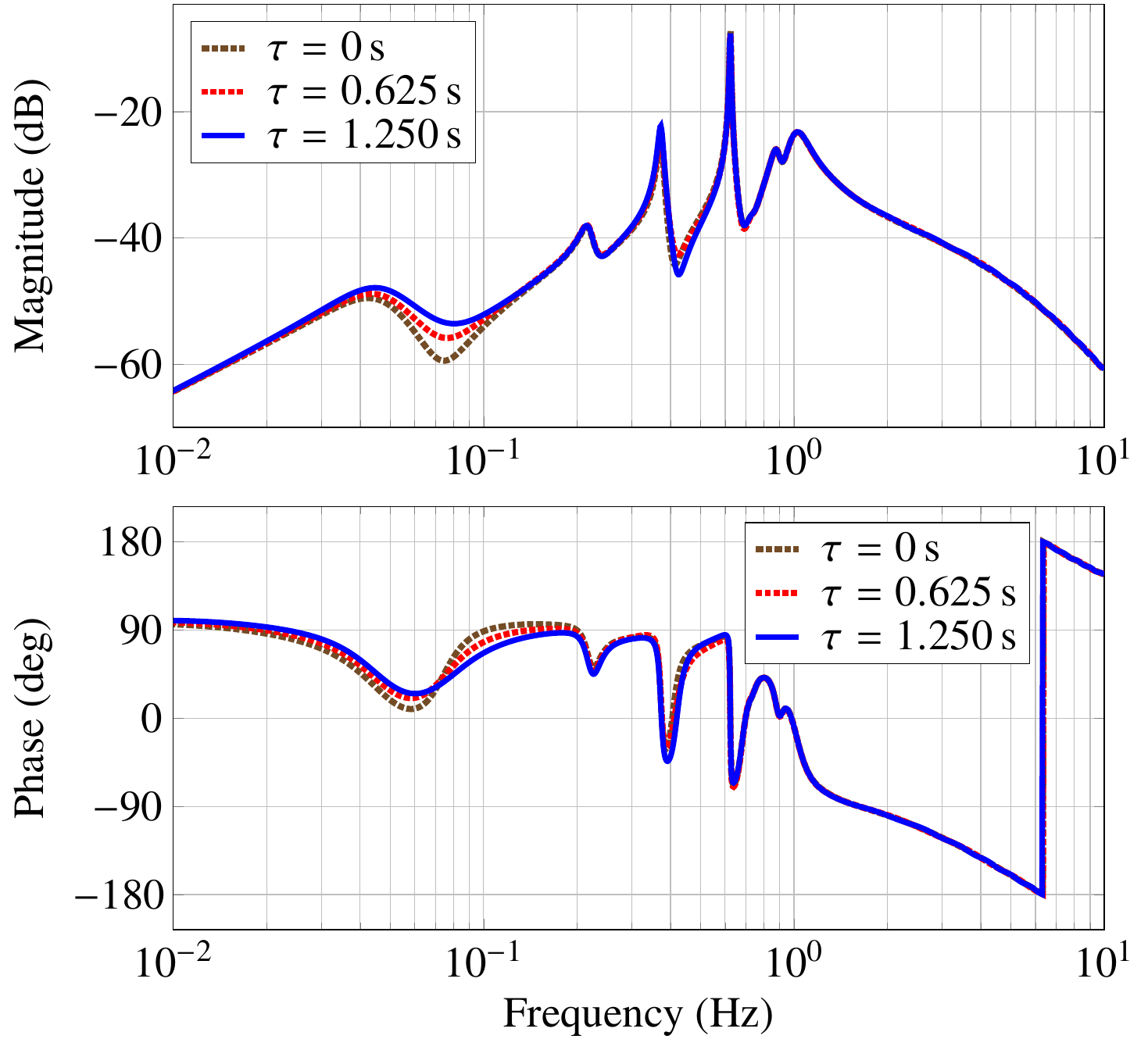}
    \caption{The effect of the combined delay $\tau$ on the open-loop
    frequency response for generator G2 with
    ${\beta_{1} = 1}$ and ${\beta_{2} = 0.5}$.}
    \label{fig:delay_xfer_fun_beta1}
\end{figure}

Modifying the state-space output matrix in \cref{eq:lti_input_long} to
account for delays as in~\cite{wilches:18}, we have
\begin{align}
    \label{eq:output_matrix_delay}
    \widehat{C}_{\nu}(s) &=
        \begin{bmatrix}
            0 & \gamma_{1}e^{-s\tau_{1}} & \gamma_{2}e^{-s\tau_{2}}
            & \dots & -\beta_{1}
        \end{bmatrix} \\
    \label{eq:xfer_delay}
    \widehat{H}(s) &= \widehat{C}_{\nu}(s)\left[sI - A\right]^{-1}B_{p},
\end{align}
where $\tau_{k}$ is the delay of the $k$th sensor.  Thus, the output
matrix changes as a function of frequency.  The open-loop transfer
function with delay is given by \cref{eq:xfer_delay}.
\Cref{fig:delay_xfer_fun_beta1} shows the results of using
\cref{eq:xfer_delay} to evaluate the effect of delay on the open-loop
frequency response for generator G2 with ${\beta_{1} = 1}$ and
${\beta_{2} = 0.5}$. For simplicity, ${\tau_{k} = \tau}$ for all $k$.
The entries of \cref{eq:output_matrix_delay} correspond to the case
where the local signal is not delayed, and the local and remote
measurements are not time-aligned upon arrival.  As a result,
${\widehat{H}(s) \neq H(s)e^{-s\tau}}$.  In the extreme case where
${\tau = \SI{1.25}{\second}}$ shown in
\Cref{fig:delay_xfer_fun_beta1}, the gain and phase are altered
slightly in the neighborhood of the frequency regulation mode;
however, the control performance and stability margins are essentially
unchanged.

\begin{figure}[!t]
    \centering
    \includegraphics[width=0.395\textwidth]{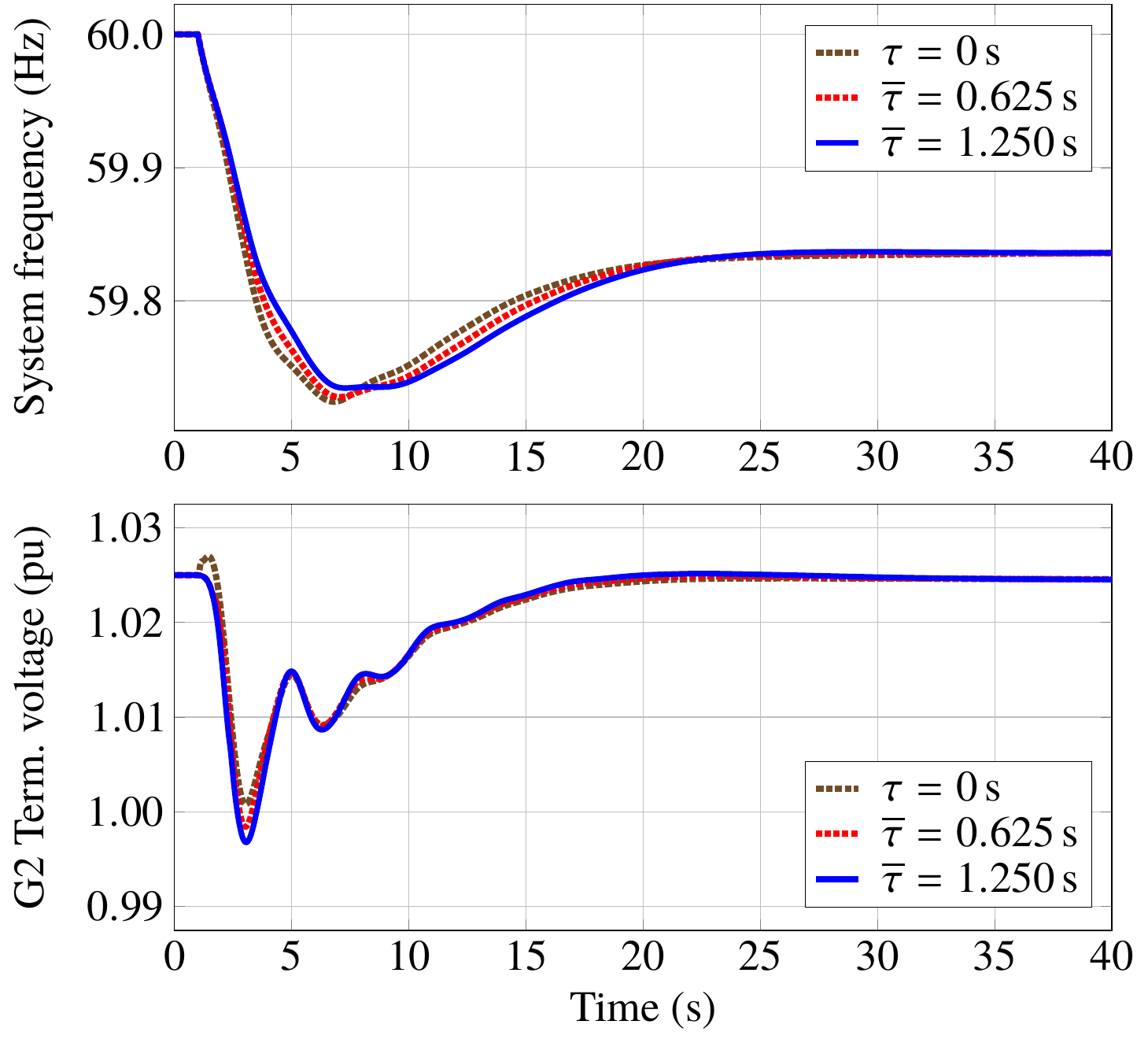}
    \caption{Simulations of generator G26 being tripped
    offline for various average combined delays
    where ${\beta_{1} = 1}$ and ${\beta_{2} = 0.5}$.}
    \label{fig:helics_time_plots_beta1}
\end{figure}

To study the impact of nonideal communication performance in the time
domain, we used a co-simulation framework called
HELICS~\cite{palmintier:17}. A communication network model for the
miniWECC was developed in \mbox{ns-3}~\cite{riley:10}. It features PMU
endpoints that communicate with the controllers via the User Datagram
Protocol (UDP).  This model includes transmission delay, congesting
traffic, and packet-based error emulation. Each generation unit in the
PST model was outfitted with a generalized $\Delta\omega$ PSS where
${\beta_{1} = 1}$, ${\beta_{2} = 0.5}$, and ${K = 9}$.
\Cref{fig:helics_time_plots_beta1} shows time-domain simulations of
generator G26, a large nuclear plant in Arizona, being tripped offline
for various expected delays $\overline{\tau}$. The results are in
close agreement with the frequency-domain analysis shown in
\cref{fig:delay_xfer_fun_beta1}.  Thus, for this example, the benefits
of the control strategy are retained even under pessimistic
assumptions of communication network performance.

In the miniWECC examples discussed herein, the center-of-inertia speed
estimate $\overline{\omega}(t)$ was synthesized using \num{30} sensors
geographically distributed throughout the system.  The effect of delay
on the open-loop frequency response is dependent on the number and
placement of the frequency (or speed) sensors.  These factors
determine how well $\overline{\omega}(t)$ tracks the target defined in
\cref{eq:coi_speed}. By extension, they also influence its spectral
content.  When $\overline{\omega}(t)$ approximately tracks the true
center-of-inertia speed, its amplitude spectrum is dominated by very
low-frequency content, generally ${\le\,\SI{0.1}{\hertz}}$. If too few
sensors are used to synthesize this estimate, and/or those sensors are
not adequately distributed, the amplitude spectrum of
$\overline{\omega}(t)$ may include significant higher-frequency
content, in and above the range of the electromechanical modes. If
this occurs, the delay may impart larger deviations in the phase
response above the frequency regulation mode than shown in
\cref{fig:delay_xfer_fun_beta1}.  For similar reasons, the
coefficients of the linear combination in \cref{eq:coi_freq} also
affect the relationship between the combined delay and the frequency
response.

The other main factor influencing this relationship is the tuning
determined by ${\beta_{1},\,\beta_{2}}$.  Analysis indicates that tunings
where ${\beta_{1} < \beta_{2}}$ may be more susceptible to the effects
of delay than those where ${\beta_{1} \ge \beta_{2}}$.  To explore
this behavior, we will analyze the entries of the output matrix
$\widehat{C}_{\nu}(s)$. Let
${\widehat{\gamma}_{k} = \gamma_{k}/\beta_{1}}$.  The matrix
$\widehat{C}_{\nu}(s)$ may then be expressed as
\begin{equation}
    \label{eq:out_mat_scaled}
    \widehat{C}_{\nu}(s) =
    \beta_{1}
    \begin{bmatrix}
        0 & \widehat{\gamma}_{1}e^{-s\tau_{1}} &
        \widehat{\gamma}_{2}e^{-s\tau_{2}} & \dots & -1
    \end{bmatrix}
    ,
\end{equation}
where
\begin{equation}
    \label{eq:gamma_hat}
    \widehat{\gamma}_{k} = \frac{\alpha_{k}}{f_0}\left(1-\frac{\beta_{2}}{\beta_{1}}\right).
\end{equation}
Recall from \cref{eq:coi_freq} that the weights $\alpha_{k}$ are
nonnegative and sum to one. For all real $\omega\tau$, it holds that
${\lvert e^{-j\omega\tau}\rvert = 1}$.  Thus, the relationship between
the magnitudes of the entries of $\widehat{C}_{\nu}(s)$ corresponding
to the delayed and non-delayed system states is primarily determined
by the ratio $\beta_{2}/\beta_{1}$.  \Cref{tab:beta_ratio} shows a
breakdown of the possible cases.  The term inside the brackets of
\cref{eq:out_mat_scaled} corresponding to the local rotor speed always
has a magnitude of one.  The magnitudes of the remaining entries
may either be zero, bounded, or unbounded.

\begin{table}
    \renewcommand{\arraystretch}{1.15}
    \centering
    \caption{Effect of Control Parameters on $\widehat{\gamma}_{k}$ Coefficients}
    \label{tab:beta_ratio}
    \begin{tabular}{l r}
        \toprule
        Parameter Ratio & Coefficient Range \\
        \midrule
        $\begin{aligned}[t]
        \beta_{2}/\beta_{1} &< 1 \\
        \beta_{2}/\beta_{1} &= 1 \\
        \beta_{2}/\beta_{1} &> 1 \\
        \end{aligned}$
        &
        $\begin{aligned}[t]
        0 &< \widehat{\gamma}_{k} \le \alpha_{k}/f_{0} \\
        0 &\le \widehat{\gamma}_{k} \le 0 \\
        -\infty &\le \widehat{\gamma}_{k} < 0 \\
        \end{aligned}$ \\
        \bottomrule
    \end{tabular}
\end{table}

When ${\beta_{2}/\beta_{1} = 1}$ the controller is immune to delay
because ${\widehat{\gamma}_{k} = 0}$ for all $k$.  This aligns with
expectations because the case where ${\beta_{1} = \beta_{2} \neq 0}$
corresponds to a standard $\Delta\omega$ stabilizer, as shown in
\cref{tab:beta_breakdown}.  When ${\beta_{2}/\beta_{1} < 1}$, the
magnitude of $\widehat{\gamma}_{k}$ has an upper bound of
$\alpha_{k}/f_{0}$.  This corresponds to the case where
${\beta_{1} > \beta_{2}}$, and the PSS prioritizes the damping of
local and inter-area modes. When ${\beta_{2}/\beta_{1} > 1}$,
$\widehat{\gamma}_{k}$ is unbounded below. Thus, the magnitude of
$\widehat{\gamma}_{k}$ may grow arbitrarily large as
${\beta_{1}\rightarrow 0}$. This does not imply that
$\widehat{C}_{\nu}(s)$ may have infinite values; rather, that the
steady-state component of the control error \cref{eq:control_error}
may be much larger than the small-signal component. This corresponds
to the case where ${\beta_{1} < \beta_{2}}$, and the PSS prioritizes
shaping the system response to transient disturbances.  It is observed
that the sensitivity of the open-loop frequency response to delay
increases as the ratio $\beta_{2}/\beta_{1}$ increases.  We
hypothesize that the driving factor in this relationship is that as
$\beta_{2}/\beta_{1}$ grows, so too do the magnitudes of the entries
of $\widehat{C}_{\nu}(s)$ corresponding to the delayed system states
in relation to the non-delayed state(s).  That is, when
${\beta_{1}\ll\beta_{2}}$, it follows that
${\lvert\widehat{\gamma}_{k}\rvert > 1}$ for some $k$.  This suggests
that if ${\beta_{1} < \beta_{2}}$, the ratio $\beta_{2}/\beta_{1}$
should be kept small.

\begin{figure}[!t]
    \centering
    \includegraphics[width=0.395\textwidth]{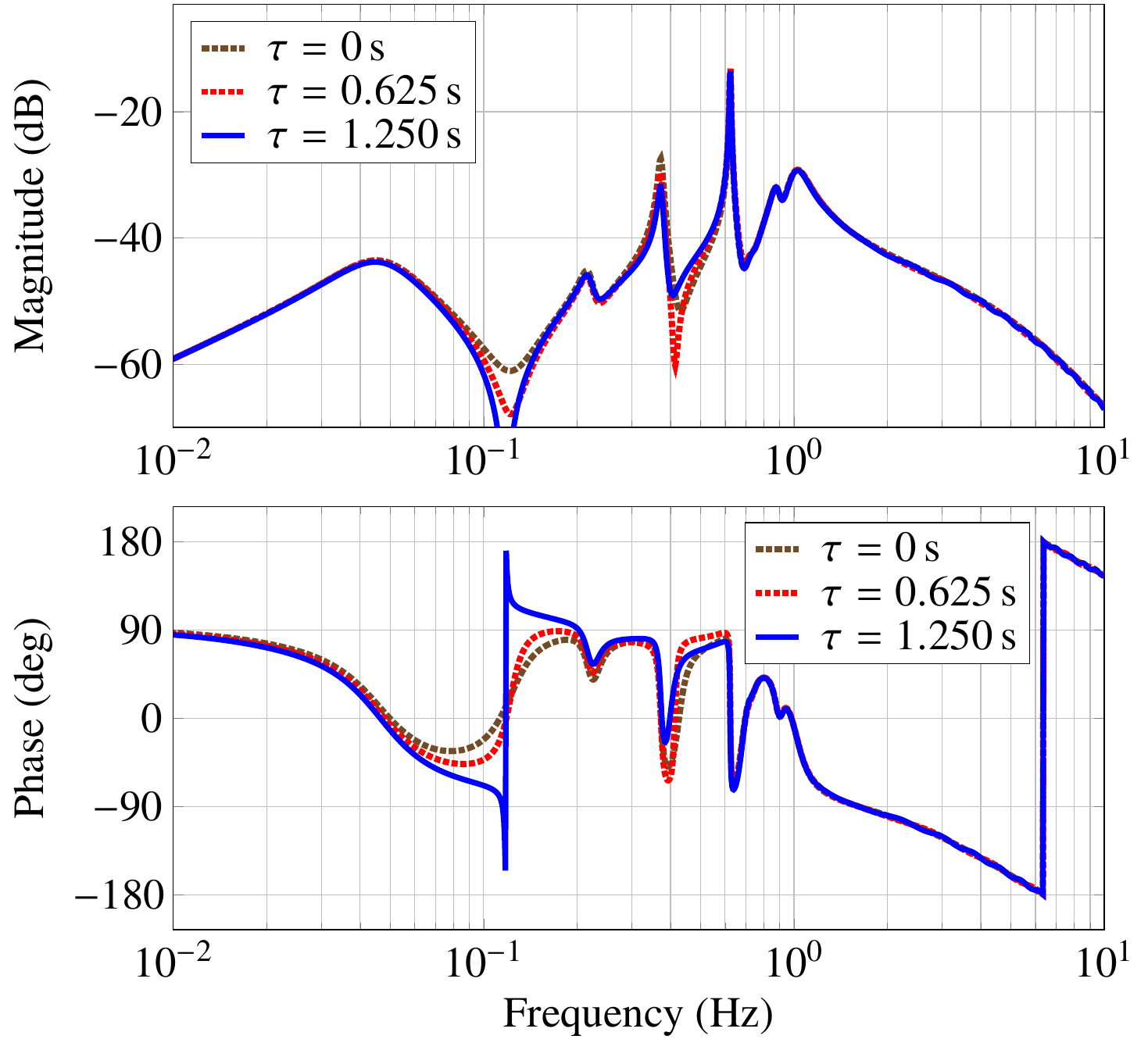}
    \caption{The effect of the combined delay $\tau$ on the open-loop
    frequency response for generator G2 with
    ${\beta_{1} = 0.5}$ and ${\beta_{2} = 1}$.}
    \label{fig:delay_xfer_fun_beta2}
\end{figure}

To illustrate this behavior, suppose that ${\beta_{1} = 0.5}$,
${\beta_{2} = 1}$ for generator G2, where ${\beta_{2}/\beta_{1} = 2}$.
\Cref{fig:delay_xfer_fun_beta2} shows the effect of delay on the
open-loop frequency response for G2 in this case.  As the delay
increases, a key transfer function zero changes position in the
complex plane. \Cref{fig:delay_xfer_fun_beta2} indicates that this
zero is pushed across the $j\omega$-axis between
\num{0.1}--\SI{0.2}{\hertz} and into the right half of the complex
plane as the delay increases.  Right-half-plane zeros, especially in
the neighborhood of the electromechanical modes, may erode stability
margins and are generally undesirable~\cite{chow:00}.  In this case,
the system remains stable when ${\tau = \SI{1.25}{\second}}$ because
the gain at the critical frequencies is very low.

\begin{figure}[!t]
    \centering
    \includegraphics[width=0.395\textwidth]{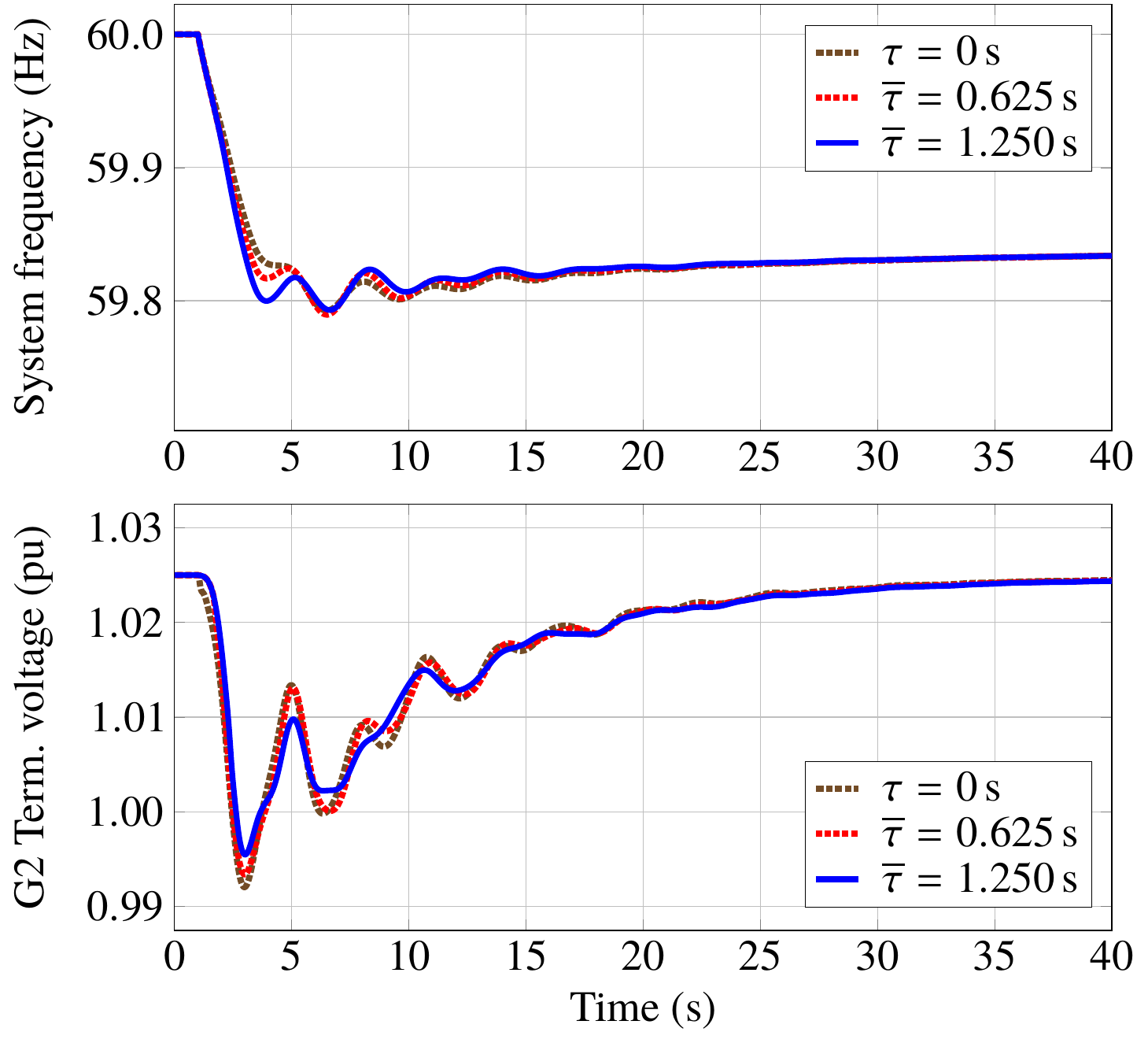}
    \caption{Simulations of generator G26 being tripped
    offline for various average combined delays
    where ${\beta_{1} = 0.5}$ and ${\beta_{2} = 1}$.}
    \label{fig:helics_time_plots_beta2}
\end{figure}

\Cref{fig:helics_time_plots_beta2} shows the system response in the
time domain following a trip of G26.  Each generation unit in the PST
model was outfitted with a generalized $\Delta\omega$ PSS where
${\beta_{1} = 0.5}$, ${\beta_{2} = 1}$, and ${K = 9}$.  As the state
trajectories show, this tuning places more emphasis on shaping the
transient response than on damping local and inter-area modes.  Such
parameter combinations should be used with caution.  Careful stability
analysis must be carried out to ensure that it is safe to employ a
particular tuning given the performance characteristics of the
measurement, communication, and control equipment.

\section{Conclusion}
\label{sec:conclusion}
This paper presented a generalization of the standard
$\Delta\omega$-type stabilizer.  It works by incorporating local
information with a real-time estimate of the center-of-inertia speed.
The ability of the stabilizer to improve the damping of
electromechanical modes is decoupled from its role in shaping the
system response to transient disturbances.  Hence, the interaction
between the PSS and AVR can be fine-tuned based on voltage
requirements.  Future work will explore a variation of the proposed
architecture that permits integral of accelerating power feedback for
mitigating torsional oscillations.  Finally, another interesting
avenue of research will be developing online methods to optimally
estimate the center-of-inertia frequency in the presence of delays,
jitter, and measurement noise.

\ifCLASSOPTIONcaptionsoff
  \newpage
\fi

\IEEEtriggeratref{45}


\bibliographystyle{IEEEtran}
\bibliography{IEEEabrv,./bib/ltv_pss_paper_arxiv}

%
%




\vskip 4pt
\vskip 1\baselineskip plus -1fil
\begin{IEEEbiographynophoto}{Ryan Elliott} received the M.S.E.E.\
degree in 2012 from the University of Washington, Seattle, WA, USA,
where he is currently a Ph.D.\ candidate in the Department of Electrical
and Computer Engineering. His research focuses on renewable energy
integration, wide-area measurement systems, and power system operation
and control. From 2012 to 2015, he was with the Electric Power Systems
Research Department at Sandia National Laboratories. While at Sandia,
he served on the WECC Renewable Energy Modeling Task Force, leading
the development of the WECC model validation guideline for
central-station PV plants. In 2017, he earned an R\&D~100 Award for
his contributions to a real-time damping control system using PMU
feedback.
\end{IEEEbiographynophoto}

\vskip 0pt plus -1fil
\begin{IEEEbiographynophoto}{Payman Arabshahi} received the
Ph.D. degree in 1994 from the University of Washington, Seattle, WA,
USA, where he is currently an Associate Professor of Electrical and
Computer Engineering and a principal research scientist with the
Applied Physics Laboratory. His research focuses on wireless
communications and networking, sensor networks, signal processing,
data mining and search, and biologically inspired systems. From 1994
to 1996, he served on the faculty of the Electrical and Computer
Engineering Department, University of Alabama, Huntsville, AL, USA.
From 1997 to 2006, he was on the senior technical staff of NASA's Jet
Propulsion Laboratory in the Communications Architectures and
Research Section. While at JPL he also served as affiliate graduate
faculty at the Department of Electrical Engineering, California
Institute of Technology, Pasadena, CA, USA, where he taught the
three-course graduate sequence on digital communications.
\end{IEEEbiographynophoto}

\vskip 0pt plus -1fil
\begin{IEEEbiographynophoto}{Daniel Kirschen} received the
Electro-Mechanical Engineering degree from the Free University of
Brussels, Belgium and the Ph.D. degree from the University of
Wisconsin, Madison, WI, USA. He is currently the Donald W. and Ruth
Mary Close Professor of Electrical and Computer Engineering at the
University of Washington, Seattle, WA, USA. His research focuses on
the integration of renewable energy sources in the grid, power system
economics, and power system resilience. Prior to joining the
University of Washington, he taught for 16 years at the University of
Manchester, U.K. Before becoming an academic, he worked for Control
Data and Siemens on the development of application software for
utility control centers. He has co-authored two books on power system
economics and reliability standards for electricity networks.
\end{IEEEbiographynophoto}


%
%




\end{document}